\documentclass[prb,reprint,aps,floatfix,showpacs,superscriptaddress,longbibliography]{revtex4-1}
\usepackage{amssymb}
\usepackage{bbm}

\usepackage{amsmath}
\usepackage{graphicx}
\usepackage[caption=false]{subfig}
\usepackage[colorlinks=true,linkcolor=blue,anchorcolor=red,citecolor=blue,urlcolor=blue]{hyperref}
\begin{document}
	\title{Universal finite size scaling around tricriticality between topologically ordered, SPT, and trivial phases}
	\author{Ke Wang}
	\email{kewang@umass.edu}
	\author{T. A. Sedrakyan}%
	\email{tsedrakyan@physics.umass.edu}
	\affiliation{%
		Physics Department, University of Massachusetts, Amherst, Massachusetts 01003, USA\\
	}%
	\date{\today}
	\begin{abstract} 
	    A quantum tricritical point is shown to exist in coupled time-reversal symmetry (TRS) broken Majorana chains. The tricriticality separates topologically ordered, symmetry protected topological (SPT), and trivial phases of the system.
        Here we demonstrate that the breaking of the TRS manifests itself in the emergence of a new dimensionless scale, 
        $g = \alpha(\xi) B \sqrt{N}$, where $N$ is the system size, $B$ is a generic TRS breaking field, and $\alpha(\xi)$, with $\alpha(0)\equiv 1$, is a model-dependent function of the localization length, $\xi$, of boundary Majorana zero modes at the tricriticality. This scale determines the scaling of the finite size corrections around the tricriticality, which are shown to be {\it universal}, and independent of the nature of the breaking of the TRS. 
        We show that the single variable scaling function, $f(w)$, $w\propto m N$, where $m$ is the excitation gap, that defines finite-size corrections to the ground state energy of the system around topological phase transition at $B=0$, becomes double-scaling, $f=f(w,g)$, at finite $B$. We realize TRS breaking through three different methods with completely different lattice details and find the same universal behavior of $f(w,g)$. In the critical regime, $m=0$,  the function $f(0,g)$ is nonmonotonic, and reproduces the Ising conformal field theory scaling only in limits $g=0$ and $g\rightarrow \infty$. The obtained result sets a scale $N \gg 1/(\alpha B)^2$ for the system to reach the thermodynamic limit in the presence of the TRS breaking.   
        We derive the effective low-energy theory describing the tricriticality and analytically find the asymptotic behavior of the finite-size scaling function. Our results show that the boundary entropy around the tricriticality is also a universal function of $g$ at $m=0$.

    \end{abstract}
    \maketitle
    \section{introduction}
    Finite-size corrections to the ground state energy in $1+1$D conformal field theories\cite{npb84VBelavin,prl86Cardy,prl86Affleck} (CFT) have been shown to be universal and were studied in various systems since mid 1980s. Recently, 
    in Ref.~\onlinecite{prl16Gulden}, it has been shown that the finite size corrections to the ground-state energy across topological phase transitions can differentiate between different topological and topologically trivial phases. These include transitions within phases classified by the group $Z$ of topological invariants (from phase characterized by topological index $n$ to the phase characterized by index $(n-1)$), or transitions from phases with the $\mathbb{Z}$ indices and phases with $\mathbb{Z}_2$ indices. Importantly, it has been shown that the finite-size scaling function is universal for all five topological symmetry classes in 1+1D (AIII, BDI, CII, D, DIII), tabulated according to Cartan's classification of symmetric spaces\cite{prb97Altland,pu01Kitaev,prb08Schnyder,rmp16Chiu}

    Although the fact of distinct  universality of the scaling function across a continuous phase transition between phases characterized by $\mathbb{Z}$ and/or $\mathbb{Z}_2$ index classification is amazing, the question arises whether these scaling properties survive in the presence of tricriticality.
        Different scaling properties can be expected for example when there is a tricritical point in the phase diagram separating three different phases: a topologically ordered phase\cite{jop15wen,aop14Greiter}, a time-reversal symmetry protected topological (SPT) phase\cite{prb13Chen,prb11pollmann,prb11Fidkowski,prb12pollaman,prb10Fidkowski,prl05Kane,np09Moore,arcmp15Senthil}, and a trivial phase. 
        Such a situation shows up if the transition between phases is accompanied by the time-reversal symmetry (TRS) breaking. Examples include transitions between topological phases belonging to the 
    BDI class classified via $\mathbb{Z}$ index and the TRS broken phase belonging to the D class classified via $\mathbb{Z}_2$ index.
    
    In this paper, we answer this question through confining our focus on  
    coupled 1+1D Kitaev-Majorana superconducting wires \cite{aop61Lieb,np15Sarma,prl10Lutchyn,prb14Wakatsuiki,pla16Wang,ropip12Alicea,iop12Leijnse,arcmpBeenakker} (throughout this article referred to as Majorana chains) from the abovementioned symmetry classes. The transition from a BDI phase to a D phase is characterized by a generic TRS breaking field, which we denote by B and which will be specified for all models considered in this work.  
The universal properties around tricriticality between BDI, D, and trivial phases are described by the
low-energy excitations around the Fermi surface, which we will discuss here in detail. 
    
    For a 1+1D critical quantum system, the CFT predicts a universal finite size scaling of the ground state energy with open boundary conditions,\cite{prl86Affleck}
    \begin{eqnarray}
    E(N)=N\epsilon+b-\frac{c}{N} \frac{\pi}{24}+\mathcal{O}(N^{-2}), \label{conformal_result}
    \end{eqnarray}
    where $c$ is the central charge\cite{npb84VBelavin,fran12conformal} of the CFT, $E(N)$ is the ground state energy of the system with size $N$, $\epsilon$ is average bulk energy, $b$ is the 
    boundary energy (for a detailed discussion of $\epsilon$ and $b$ see Ref.~\onlinecite{prl86Affleck} and the main text). 
    
    Around topological quantum phase transition in $1+1$D, the CFT result  Eq.(\ref{conformal_result}) is generalized to\cite{prl16Gulden} 
    \begin{eqnarray}
    E(N,m)=N\epsilon(m)+b(m)-\frac{c}{N} f(Nm)+\mathcal{O}(N^{-2}), \label{Alex}
    \end{eqnarray}
    where $f(Nm)$ is shown to be a universal function of its argument for five different symmetry classes. This is achieved in the double scaling limit, when $N\rightarrow \infty$, $m \rightarrow 0$, while $w=Nm$ is kept constant. 
    This result is derived from $1+1$D Majorana field theory, which describes the critical phases of free fermion models. 
    
While the finite-size scaling across the topological quantum phase transition is now well understood, we will show that the situation is different around the tricriticality, where there are two extra Majorana edge modes near the Fermi surface. 
There are two categories of states determining the low-energy sector: (i) The states described by the 1+1D massive (with the excitation gap, $m$) bulk Majorana field theory (\text{MFT}). 
(ii) A boundary Hamiltonian describing an even number of localized Majorana edge modes with localization lengths $\xi$.
Once the TRS breaking field is introduced, it can drive the boundary RG flow\cite{prl91Affleck} from the tricriticality (gapless SPT phase)\cite{Verresen_2019,prb18parker} to the criticality without edge modes (gapless trivial phase). Such non-trivial boundary effects can exhibit themselves in the scaling properties of finite size corrections to the ground state energy. This is the effect that we are going to explore here. 

    In this work, we derive the finite size scaling function around the tricriticality, which is strongly influenced  by the TRS breaking field, $B$.  We show that the result Eq.(\ref{Alex}) is generalized to  
    \begin{eqnarray}
    E=N\epsilon+b-\frac{c}{N} f(Nm,\alpha\sqrt{N}B)+\mathcal{O}(N^{-2}). \label{Ke_scaling}
    \end{eqnarray}
    where $E=E(N,m,B)$ is the ground state energy that now also depends on $B$, $c=1/2$, and $\alpha(\xi)$ is a function of the localization length of the Majorana edge modes, $\xi$, at the tricriticality. The function $\alpha(\xi)$ depends on the details of the lattice model, but $\alpha(0)\equiv 1$ for all of them. The finite size scaling function $f(w,g)$ is now a function of two variables: $w=Nm$ and $g=\alpha \sqrt{N}B$. Under two simultaneous  double-scaling\cite{prb76Wu} limits: a) $N\rightarrow\infty$, $m\rightarrow0$ with $w=const$ and b) $N\rightarrow\infty$, $B\rightarrow0$, with $g=const$, the function $f(w,g)$ is a universal function of $w$ and $g$.  This is the main result of our work, which will be obtained below both, analytically and numerically. We will also discuss the implications of the new emergent scale on the numerical simulations of many-body systems with TRS breaking. 
    
    The universality of the double-scale function $f(w,g)$ is shown for a parent Hamiltonian representing a pair of coupled Majorana chains (discussed in Sect.(\ref{full})), with three different symmetry breaking fields $B$: 
    
    (i) We consider two Majorana chains coupled to each other with a complex pairing-potential $\Delta_v=\Delta_v^R+i\Delta_v^I$,\cite{prb14Wakatsuiki} along the {\em vertical} rungs. The TRS breaking field, in this case, is identified with $B\equiv\Delta^I_v/(2t)$, where $t$ is the nearest-neighbor hopping parameter.  In this model, the function $\alpha(\xi)$ is found to be $\alpha(\xi)=\sqrt{\coth(1/2\xi)}$.
    The Hamiltonian of the model, its solution, along with the detailed analysis of the ground state energy is discussed in Sect.(\ref{comlex__pairing_potential_section}). 
    
(ii) We consider coupled Majorana chains in a uniform external magnetic field\cite{pla16Wang}.  The Flux can be realized by complex hopping $t e^{i\theta/2 }$ along the {\em horizontal} chains. The TRS breaking field in this case is identified with $B=\frac{1}{2}\sin \left( \theta/2 \right)$. 
This model is analysed in Sect.(\ref{uniform_flux_sec}). 

    (iii) We study coupled Majorana chains in the presence of a staggered magnetic flux, $\pm\theta$, threading square plaquettes of the lattice. The Flux can be realized by alternating complex hopping $t_v e^{\pm i\theta/2 }$ along the {\em vertical} rungs.  In this model, 
    $\alpha(\xi)=\sqrt{\tanh(1/2\xi)}$. The TRS breaking field in this case is identified with $B=t_v \sin \left( \theta/2 \right)/(2t)$, 
    This model is analysed in Sect.(\ref{stagger_flux_sec}).
    
\begin{figure}
        \includegraphics[scale=0.5]{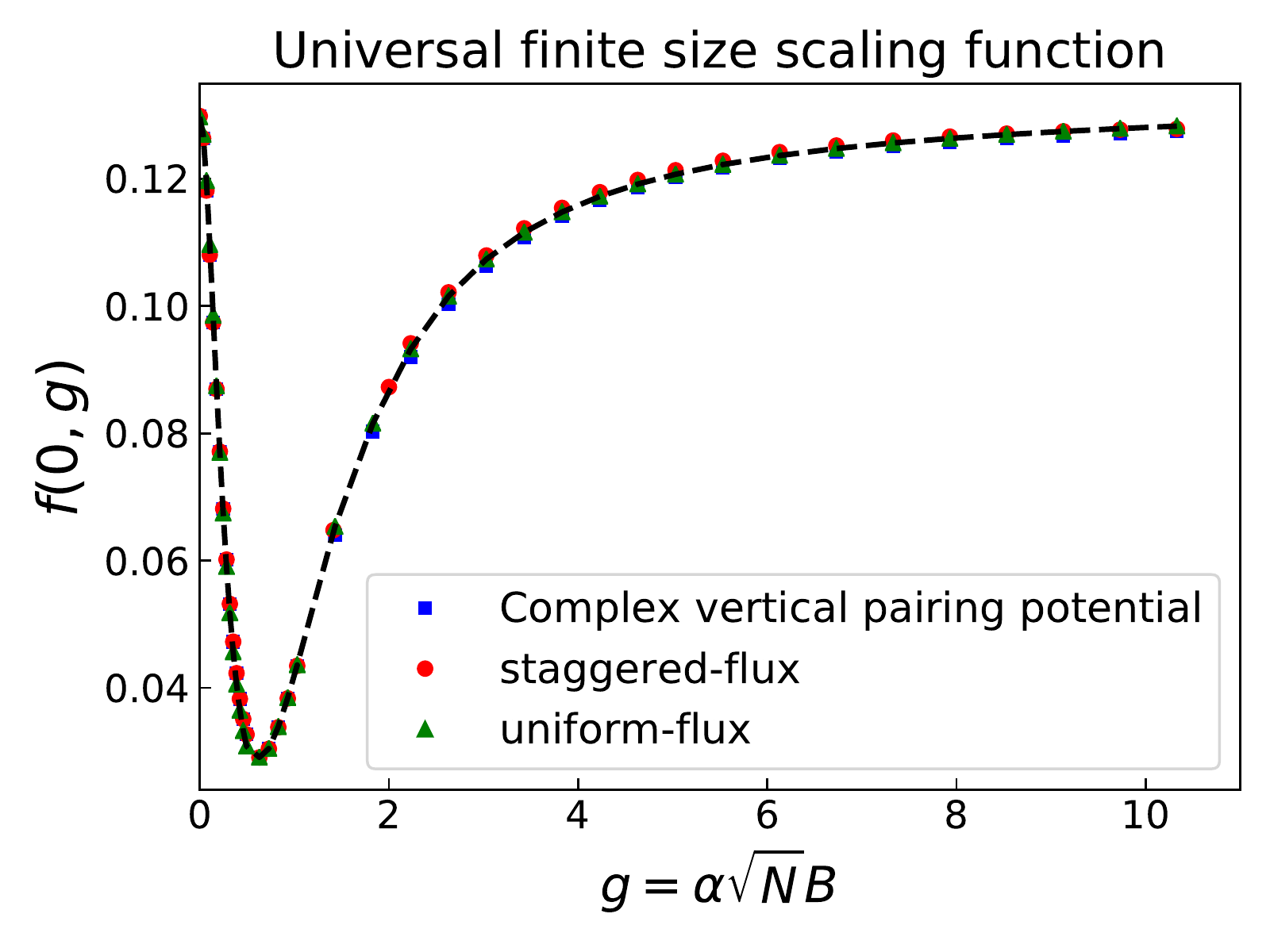}
        \centering
        \caption{(Color online) Universal finite size scaling function $f(w=0,g)$ plotted at criticality, $w=0$, versus $g$.  
        The function is nonmonotonic and exhibits a strong $g$-dependence. It starts decreasing from the Ising CFT value, $f(0,0)=\pi/24$, and undergoes a minimum at $g=0.5$. At $g\gg 1$, it gradually converges to the CFT result, $f(0,g\rightarrow \infty)=\pi/24$.
    } 
        \label{central}
    \end{figure}

\begin{figure}
        \includegraphics[scale=0.5]{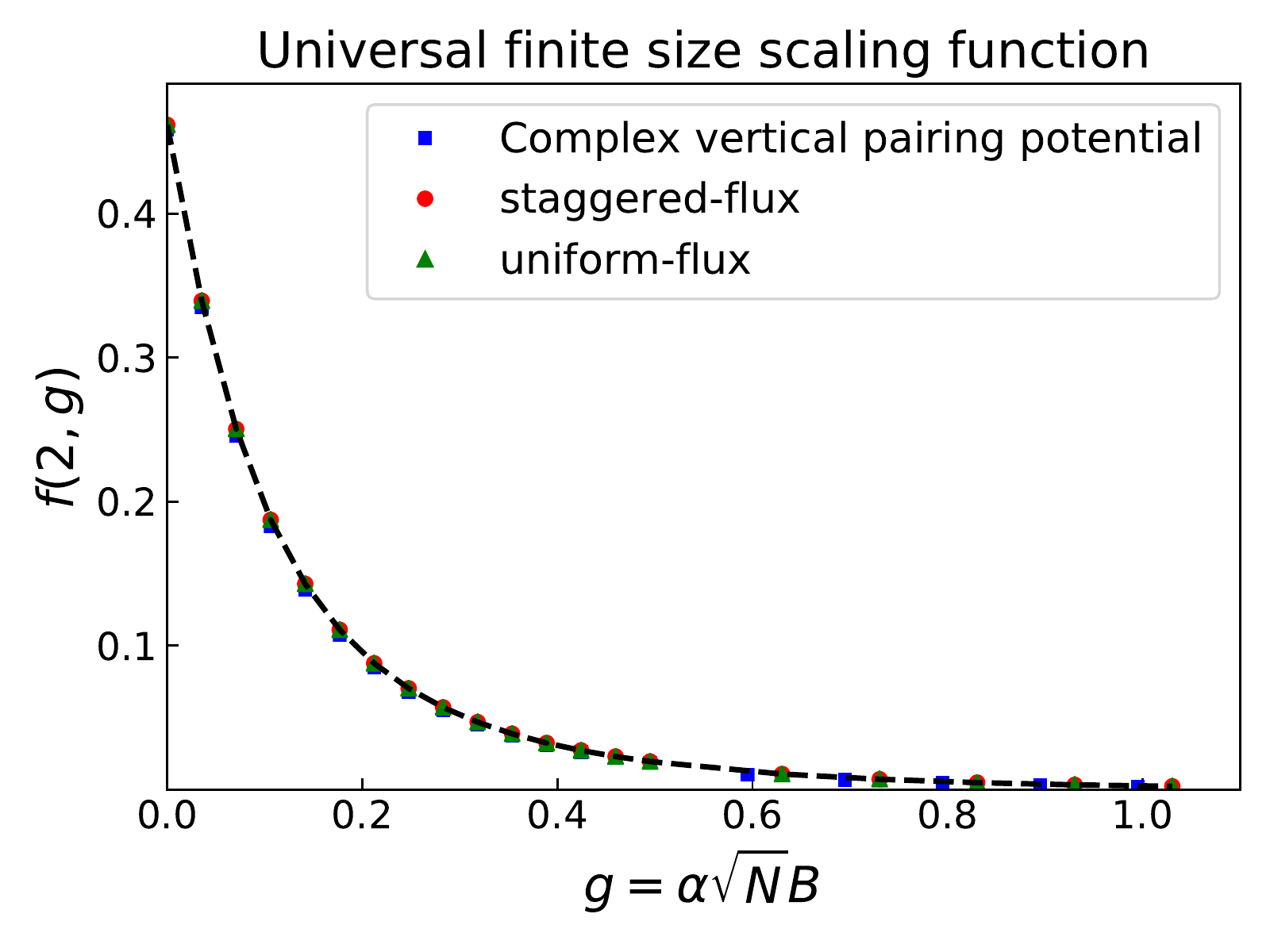}
        \centering
        \caption{(Color online) The universal finite size scaling function $f(w,g)$ is plotted vs. $g$ at $w=2$ (gapped phase) for three models with different origin of TRS breaking field, $B$. 
        } 
        \label{uw2}
    \end{figure}

    The scaling of finite-size correction to the ground state energy in all three models under consideration is shown to obey the universal behavior determined by function $f(w,g)$. Particular cases corresponding to $w=0$ (critical phase) and finite $w=2$ (gapped phase) are shown in Figs.~(\ref{central}) and (\ref{uw2}) respectively.

    Although $f(0,0)=\pi/24$, which directly follows from CFT calculations, the scaling function $f(0,g)$ exhibits nontrivial behavior.  Interestingly enough, this function, plotted in Fig.(\ref{central}),  appears to be nonmonotonic and  strongly $g$ dependent. 
    The small and large $g$ asymptotes are found to be 
    
    \begin{eqnarray}
    f(0,g)\simeq 
    \left\{
    \begin{array}{cc}
    
    \frac{\pi}{24}+\frac{1}{\sqrt{\pi}}g^2 \log g,  &\text{$g\ll 1$}\\
    \frac{\pi}{24}-\frac{\gamma}{g^2},&\text{$g\gg 1$}.\\
    \end{array}
    \right.
    \end{eqnarray}
    Here $\gamma$ is a constant, numerical value of which is found to be $\gamma\approx 0.24$.
    
    To complete the theoretical picture, we derive the low-energy effective theory describing the symmetry-breaking effect. The theory enables one to analytically uncover the scale, $g$, which describes physics around the tricriticality, and extract the asymptotic behavior of the finite size scaling function, $f$.
    We start with the full-symmetry parent Hamiltonian from BDI class and find its phase diagram in Sect.(\ref{full}).
In Sect.(\ref{breaking}), we define three different symmetry-breaking models where Majorana tricritical points emerge. In Sect.(\ref{finite}), we numerically show the emergence of $g$ in finite-size scaling function and plot $f(w,g)$ in several different cases.  
In Sections (\ref{lowformt}) and (\ref{boundary_Ising}), we derive the low-energy theory near the tricriticality and provide its solution. We also show the universal behavior of the boundary entropy in all three models.
    Conclusions are presented in Sect.(\ref{conclusion}) and some technical details are presented in the Appendices.

	 \section{Parent Hamiltonian from BDI symmetry class} \label{full}
    In this section, we write down the "parent" Hamiltonian with $T$, $P$, and $TP$ symmetries. We show that this model can support three different phases.
    These are (i) a trivial gapped phase; (ii) a topologically ordered phase supporting two boundary  Majorana modes; (iii) symmetry protected topological phase supporting four boundary Majorana modes.
    At the end of this section, we analytically derive the low-energy effective theory describing 
    the finite-size scaling effect around special critical points which belong to the XY universality class.
    
    We start with two identical Majorana chains with full symmetries of the BDI class:
    
    \begin{eqnarray}
    H&=&H_1+H_2+H_{\text{interchain}} \label{total_hamiltonian}
    \end{eqnarray}
    where $H_1$ and $H_2$ are two Hamiltonians, each representing a 1+1D chain. The  $H_{\text{interchain}}$ describes coupling between two chains that has the form of interleg hopping and pairing:
    \begin{eqnarray}
    \nonumber
    H_m&=&-\sum_{j} \mu \hat{a}^{\dagger}_{j,m}\hat{a}_{j,m}
    -\sum_{j}( t \hat{a}^{\dagger}_{j+1,m}\hat{a}_{j,m}+h.c.)\\ 
    &+&\sum_{j}(\Delta \hat{a}_{j,m}\hat{a}_{j+1,m}+h.c.), \quad m=1,2\label{parent_intra}
    \\ 
    H_{\text{interchain}}&=&\sum_j (-t_v \hat{a}^{\dagger}_{j,1}\hat{a}_{j,2}+ \Delta_v \hat{a}_{j,1}\hat{a}_{j,2}+ h.c.). \label{i}
    \end{eqnarray}
    Here $\hat{a}^{(\dagger)}_{j,m}$ are fermion annihilation (creation) operators, index $j=1\ldots L$ labels the position of a fermion, index $m$ labels the chains, $\mu$ is the chemical potential for each of the chains, $\Delta$ ($\Delta_v$) is the intrachain (interchain) pairing potential, and $t$ ($t_v$) is the intrachain (interchain) hopping parameter. 
    One can safely set pairing $\Delta$ to be real since its complex phase can be absorbed into fermion operators.
    In the presence of TRS (and obviously PHS) symmetry, $t$, $t_v$ and $\Delta_v$ are all real. 
    
    To obtain the single-particle spectrum, one may introduce the momentum-space fermion operators 
    \begin{eqnarray}
    \hat{a}_{k,m}&=&\frac{1}{\sqrt{N}} \sum_j  \hat{a}_{j,m} e^{ik j}
    \end{eqnarray} 
    where the lattice constant is set to be unity.
    Then $H$ can be represented in the momentum space as 
    \begin{eqnarray}
    H&=&\frac{1}{2}\sum_k \hat{\psi}^{\dagger}_k h(k) \hat{\psi}_k, \qquad\text{with}\\ \nonumber
    h(k)&=&(-2t \cos k-\mu)\sigma_z\mathop{\otimes} \mathbbm{1}_2+2\Delta \sin k\sigma_y \mathop{\otimes} \mathbbm{1}_2  \\
    &-&t_v\sigma_z \mathop{\otimes} \tau_x+\Delta_v \sigma_y \mathop{\otimes} \tau_y \label{BdG}
    \end{eqnarray}
    where $ \hat{\psi}^{\dagger}_k=\begin{pmatrix}
    \hat{a}^{\dagger}_{k,1}&    \hat{a}_{-k,1}&    \hat{a}^{\dagger}_{k,2}&    \hat{a}_{-k,2}
    \end{pmatrix}$, $\boldsymbol{\sigma}$ are Pauli matrices in Nambu space and $\boldsymbol{\tau}$ are Pauli matrices in the space of $m=1,2$ Majorana chains. 
    The time reversal symmetry is seen from the commutativity of the anti-unitary operator $T$ with the Hamiltonian, $T H  T^{-1} =H$, where $T$ acts on spinless fermions as $T \hat{a}_{j,i}  T^{-1} =\hat{a}_{j,i}$. 
    Similarly, the particle-hole-symmetry is implied by the operator $P=\tau_x K \mathop{\otimes} \mathbbm{1}_2  $ in BdG-formalism, where $K$ is the operator of complex conjugation. Thus, one may show that $P h(-k)  P^{-1} =- h(k) $, where $h(k)$ is the $k$-space hamiltonian in Eq.(\ref{BdG}). 
    
    Diagonalization of $h(k)$ is achieved through the Bogoliubov transformation, yielding the single-particle excitation spectrum
    \begin{eqnarray}
    E_{s}(k)&=&\sqrt{\xi_1(k)+2s\xi_2(k)} \label{energy}
    \end{eqnarray}
    where $\xi_1=(2\Delta \sin k)^2+(2t\cos k+\mu)^2+t^2_v+\Delta_v^2$, $\xi_2=\sqrt{4\Delta^2 \sin^2 k\Delta^2_v+\Delta^2_vt_v^2+(2t\cos k+\mu)^2t_v^2}$ and $s=\pm$.
    \footnote{Then one may get the ground state energy $N\epsilon$ with periodic boundary conditions
        \begin{eqnarray}
        N\epsilon&=&-\frac{N}{2} \int_{0}^{2\pi} \frac{dk}{2\pi} \sqrt{\xi_1+2s\xi_2} 
        \end{eqnarray}
        where $1/2$ comes from redundancy for degrees of freedom in BdG formalism.}
    Given the spectrum, one can obtain the phase boundaries (critical lines in the phase space of model parameters) by solving the equation, $E_{s}(k)=0$, for some $k$. This yields
    \begin{eqnarray}
    \label{yields}
    \nonumber
    (2t\cos k+\mu)^2+\Delta^2_v-t^2_v&=&0,\\
    (2t\cos k+\mu) \sin k&=&0.
    \end{eqnarray}
    One can further simplify Eqs.~(\ref{yields}) to obtain the phase boundaries (where the spectral gap closes) at $k=0;\pi$ as
    \begin{eqnarray}
    \Delta_v^2+(2t\pm\mu)^2&=&t^2_v.
    \end{eqnarray}
    The third phase boundary corresponds to momenta solving the equation $\cos k=\pm\mu/{2t}$, and parameters 
    satisfying the condition 
    \begin{eqnarray}
    t^2_v+\frac{\Delta^2}{t^2}(2t+\mu)(2t-\mu)&=&\Delta_v^2,
    \end{eqnarray}
    where $2w > \mu$. 
    
    In this work,  we are interested in characterizing the topological states via the edge Majorana modes (and thus, we do not explicitly evaluate the topological quantum numbers from the bulk wavefunctions). Since a number of stable edge modes characterize topological phases, below, we will calculate the localization lengths of Majorana modes from the lattice Hamiltonian. 
    
    Let us adopt the following notations for Majorana fermion operators
    \begin{eqnarray}
    \nonumber
    c_{2j-1,m}&=& \hat{a}_{j,m}+\hat{a}^{\dagger}_{j,m}  \\ 
    c_{2j,m}&=&\frac{ \hat{a}_{j,m}-\hat{a}^{\dagger}_{j,m}}{i} 
    \end{eqnarray}
    where anticommutator $\{c_{j,m},c_{l,n}\}=2\delta_{j,l}\delta_{m,n}$. 
    Using these notations, the Hamiltonian can thus be represented in the Majorana basis.
    The intra-chain part is given by
    \begin{eqnarray}
    H_m&=&\frac{i}{2}\sum_j\{-\mu c_{2j-1,m}c_{2j,m}+(\Delta+t)c_{2j,m}c_{2j+1,m}\nonumber \\
    &+&(\Delta-t)c_{2j-1,m}c_{2j+2,m}\}. \label{Kitaev_chain_Hamiltonian}
    \end{eqnarray}
    The interchain part is written as
    \begin{eqnarray}
    \nonumber
    H_{\text{interchain}}&=&\frac{i}{2}\sum_j(\Delta_v-t_v)c_{2j-1,1}c_{2j,2}\\
    &+&(\Delta_v+t_v)c_{2j,1}c_{2j-1,2}. \label{Inter_chain_Hamiltonian}
    \end{eqnarray}
    
\begin{figure}
        \includegraphics[scale=0.5]{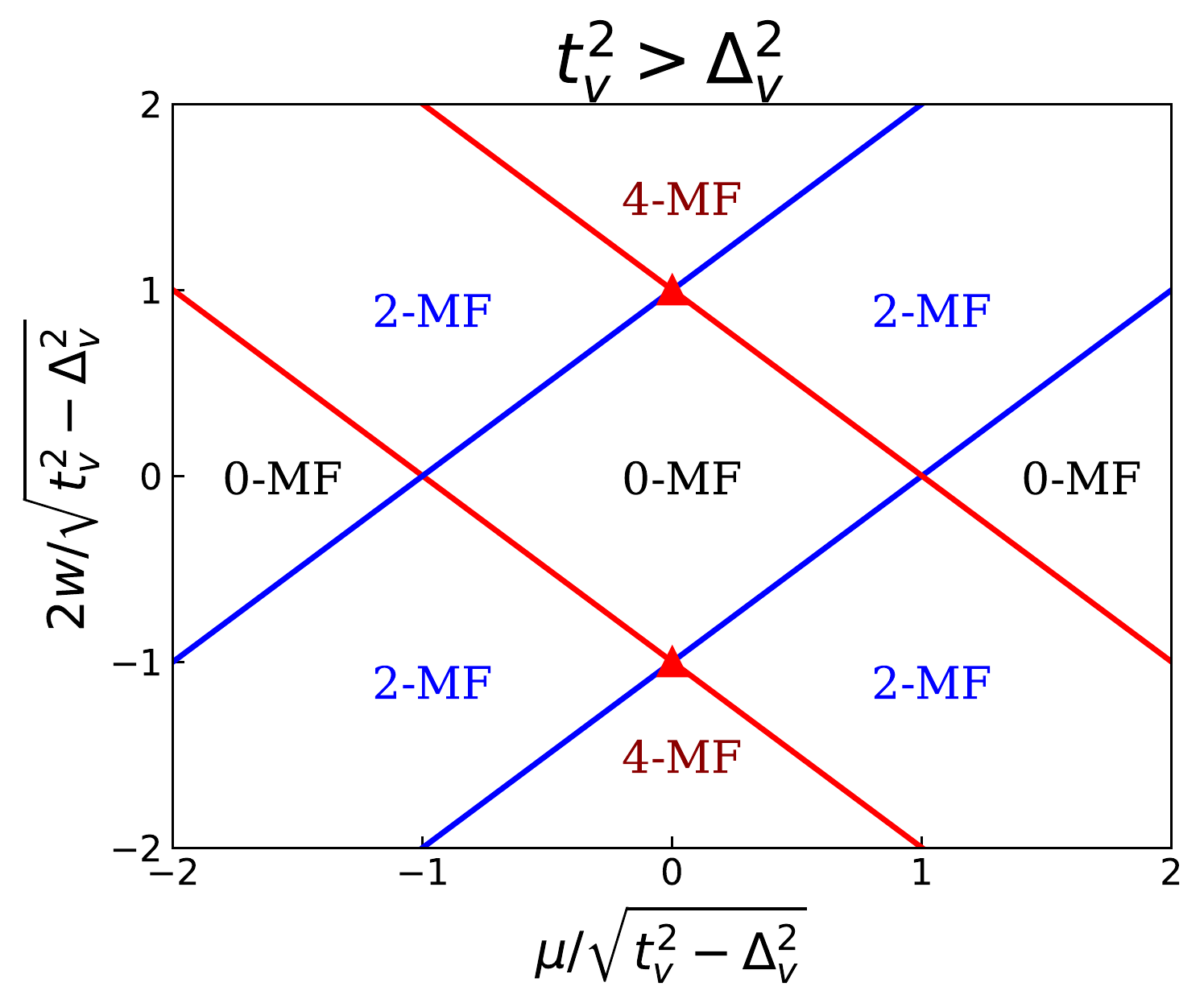}
        \centering
        \caption{(Color online) Phase Diagram for Parent Hamiltonian from BDI symmetry class. $n$-MF represents gapped phase with $n$ localized zero-energy Majorana modes. There are three different phases $0$-MF, $2$-MF and $4$-MF.
            Critical points in the phase diagram are Ising universality with central charge $c=1/2$, but intersection points of two critical lines are XY-universality with central charge $c=1$.
            Two special intersection points located at $(0,\pm 1)$ are emphasized as red triangle points. Two relevant perturbations $\mu$ and $2w$ will drive the system away from the criticality: tuning $\mu$ opens a gap to $2$-MF; tuning $2w$  opens a gap to $4$-MF or $0$-MF. 
        }
        \label{Tcphase}
    \end{figure}

    Using the Majorana representation of the Hamiltonian, we can diagonalize it and find the Majorana zero energy states that are localized at the boundaries of the chain. We show in  Appendix~\ref{Majorana_zero_modes} that the question of the existence of Majorana zero-energy states reduces to the estimation of the localization length $\xi_\pm$. The latter is found from the wavefunction that decays into the bulk exponentially as $\exp \{-x/\xi_\pm\}$, with
    \begin{eqnarray}
    \xi^{-1}_{\pm}=\ln \frac{2t}{\mu\pm \sqrt{t^2_v-\Delta^2_v}}. \label{correlation_length}
    \end{eqnarray}
    Here, if the argument of the logarithm is negative, $\xi^{-1}_{\pm}$ obtains a complex phase, $i\pi$, indicating an oscillatory wavefunction. Thus,
    \begin{itemize}
        \item (i) if $\xi_+>0$ and $\xi_->0$, the system is in the gapped phase with four localized zero-energy Majorana modes.  

\item     (ii) if $\xi_+>0$($\xi_+<0$) and $\xi_-<0$($\xi_->0$), the system is in the gapped phase with two localized zero-energy Majorana modes. 
    
    \item (iii) if $\xi_+<0$ and $\xi_-<0$, the system is in the trivially gapped phase with no localized zero-energy Majorana modes.
    
    \end{itemize}
    These three different phases are shown on the phase diagram Fig.~(\ref{Tcphase}) in the space of rescaled energies $w$ and $\mu$. 
    Throughout this paper, we will adopt the notation "$n$-MF" 
    to represent the gapped phase with $n$ localized zero-energy Majorana modes (below we will deal with cases with $n=0,2,4$).
    
    The critical lines (or, in other words, the phase boundaries)  in the phase diagram Fig.~(\ref{Tcphase}) belong to the Ising universality class with central charge $c=1/2$. The exceptions are the intersection points of two critical lines, which belong to the $XY$ universality with central charge $c=1$.  Two relevant perturbations, given by the change of critical $\mu$ and $2w$, will drive the system away from criticality. Tuning of parameter $\mu$ opens a gap and drives the system into $2$-MF state. Tuning of $2w$  opens a gap, and the system finds itself either in $4$-MF or $0$-MF state. 
    
    One can calculate the finite size corrections to the ground state energy
    around the $XY$ criticality. To this end, we define two masses $m_\pm$
    \begin{eqnarray}
    m_\pm=\frac{2t\pm\mu-\sqrt{t^2_v-\Delta^2_v}}{2t}. \label{mass_xy}
    \end{eqnarray}
     The magnitudes of $m_+$ and $m_-$ are the spectral gaps measured at $k=0$ and $k=\pi$ respectively. Interestingly, the low-energy effective theory around the criticality is given by a direct sum of two massive Majorana field theories:
    \begin{eqnarray}
    H_{\text{low}}&=&\sum_{s=\pm}\frac{iv_F}{2}\int dx \eta^T_s \begin{pmatrix}
    \partial&    m_s\\
    -m_s&    -\partial
    \end{pmatrix} \eta_s, \label{low_energy_xy}
    \end{eqnarray}
    where $\eta_\pm$ is a two-component Majorana field operator.
    To evaluate the finite-size scaling function corresponding to a single copy of Majorana field from Eq.~(\ref{low_energy_xy}), 
    one may double the number of degrees of freedom in Majorana field theory above and form a $1+1$D massive Dirac field. Then one will recover the finite-size scaling function, $f_D(w)$ (defined in Eq.~(\ref{Alex})), for $1+1$D Dirac field theory found in Ref.~\onlinecite{prl16Gulden}. 
    
    In the present situation, for low-energy Hamiltonian Eq.~(\ref{low_energy_xy}), we obtain that finite size scaling function $\tilde{f}$ becomes function of two masses, $w_+=Nm_+$ and $w_-=Nm_-$:
    \begin{eqnarray}
    \nonumber
    \tilde{f}(w_+,w_-)&=&\frac{1}{2}\{f_\text{D}(w_+)+f_\text{D}(w_-)\} \label{finite_xy},
    \end{eqnarray}  
    where the factor $1/2$ eliminates the double counting of degrees of freedom in Dirac field theory, as the latter is equivalent to the direct sum of two Majorana field theories. 
    
    Instead of plotting $\tilde{f}(w_+,w_-)$ in a 3D space as a 2D surface parametrized by two scales $w_+, w_-$, we chose $|w_+|=|w_-|=w$ and plot $\tilde{f}(w,w)$ vs $w$. This plot will bear all the necessary information about the naure of the phases as follows: 
    \begin{itemize}
        \item  i) From definition of masses in Eq.(\ref{mass_xy}), we see that if $(w_+,w_-)=(w,w)$ with $w>0$, then both masses are positive and the system is in the $4$-MF phase. Below we will use the notation $f_4(w)\equiv \tilde{f}(w,w)$, for the finite size scaling function in the $4$-MF phase. \label{4_item}
        
        \item ii) If $(w_+,w_-)=(w,-w)$ (or $(-w,w)$), $w>0$, then one of the masses is positive (indicating the existence of two localized Majorana modes)while the other is negative (indicating a trivial boundary).
        In this case one has the $2$-MF phase.  Below we will use the notation $f_2(w)\equiv \tilde{f}(w,-w)=\tilde{f}(-w,w)$ for the finite size scaling function in $2$-MF phase.  \label{2_item}
    \item  iii) Finally, when $(w_+,w_-)=(-w,-w)$, $w>0$, one obtains a trivial $0$-MF phase. The corresponding  finite size scaling function is denoted as $f_0(w)\equiv \tilde{f}(-w,-w)$.  \label{0_item}

    \end{itemize}
     
\begin{figure}
        \includegraphics[scale=0.5]{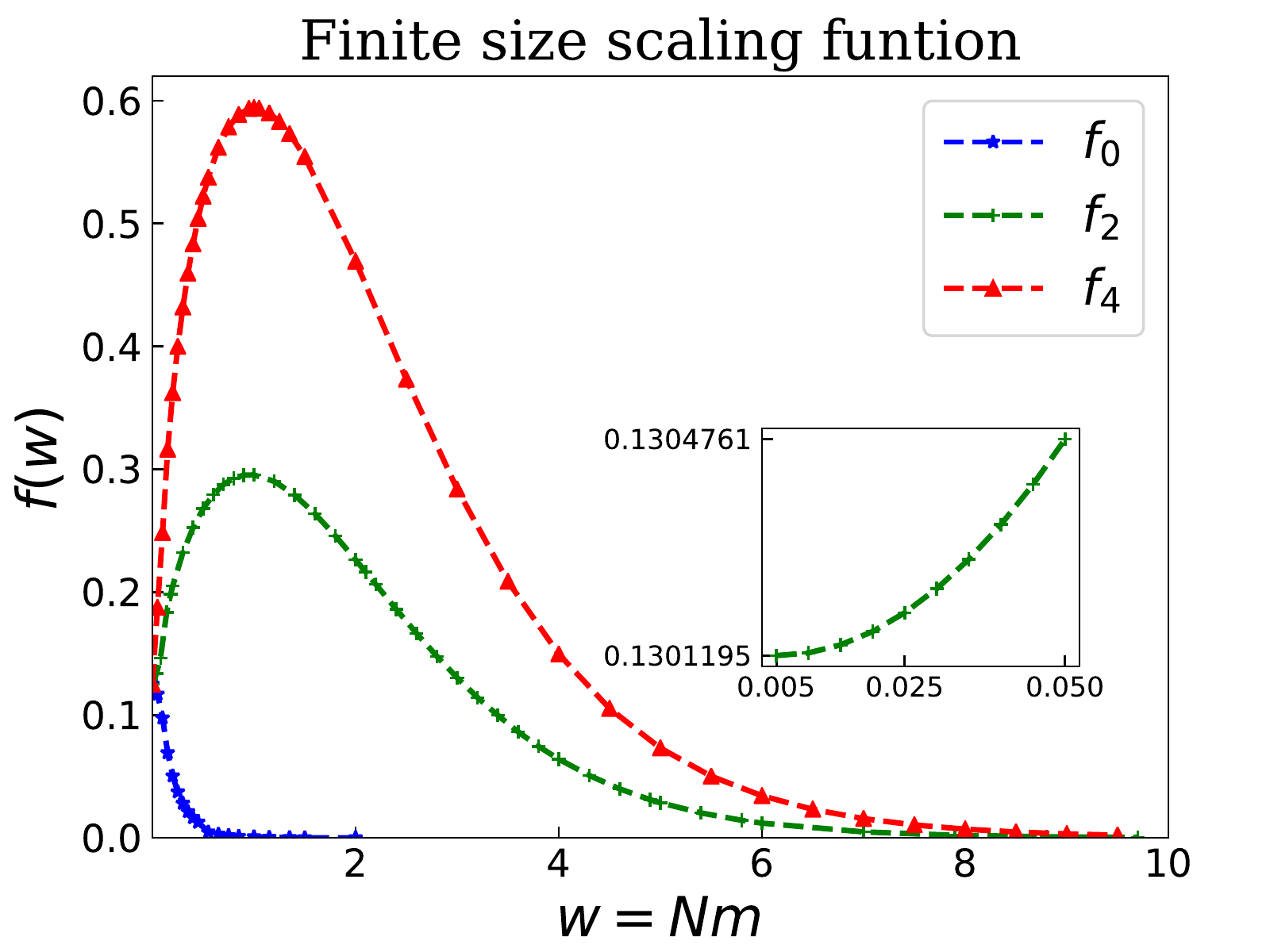}
        \centering
        \caption{(Color online) Finite size scaling functions, $f_n(w)$, for $n=0, 2, 4$ around the criticality marked by the (red) triangle in Fig.\ref{Tcphase}. Inset: The behavior of $f_2$ at small $w\ll 1$. The finite size scaling function clearly distinguishes all three different phases around the criticality.
        }
        \label{Dirac}
    \end{figure}

    Thus we obtain three finite size scaling functions, $f_0$, $f_2$ and $f_4$, of a single variable $w$. These functions are depicted in Fig.~(\ref{Dirac}). We see that $f_4$ is has a pronounced maximum at the topological $4$-MF phase; $f_2$ exhibits a less pronounced maximum in the $2$-MF phase; and $f_0$ is a featureless decaying function in the trivial $0$-MF phase. The inset to Fig.~(\ref{Dirac}) shows the small $w$ behavior of $f_2$ suggesting that it is completely regular and $df_2/dw|_{w=0}=0$. The latter property is expected from Eq.(\ref{finite_xy}). 

    Properties of $\tilde{f}(w_+,w_-)$ follow from the fact that we have a direct sum of noninteracting Hamiltonians at intersections of two critical lines in Fig.~\ref{Tcphase}. The same procedure is straightforwardly generalized to more  complex situations, where one has $N$ intersecting critical lines, each represented by a central charge $c_i$, $i=1\ldots N$, (e.g., if there are many coupled Majorana chains). In this case the finite size scaling function becomes multi-variable, and equal to $\tilde{f}(w_1,w_2,...,w_n)=\sum_{i=1}^{N} c_i f_D(w_i)/\sum_i^N c_i$.

 \section{Models with broken time-reversal symmetry
    } \label{breaking}
 In the above section, we discussed and solved models whose parameters are real, and whose Hamiltonian operators preserve time-reversal symmetry (TRS). In this section, we will consider two-leg ladder models, where the TRS is explicitly broken. This can be achieved, e.g., through endowing a \textit{complex} phase to model parameters in such a way that it cannot be gauged out. We introduce three different TRS breaking models and prove the existence of Majorana tricriticality between topologically ordered, SPT, and trivial phases in each of them. Consequently, we trace the evolution of the single-particle spectrum with varying the TRS breaking field, $B$, around the tricriticality.
    
    \subsection{Model I: Majorana ladder with complex vertical pairing potential} \label{comlex__pairing_potential_section}
    The complex phase of the pairing potential, $\Delta$, in the single Majorana chain, can be gauged out (through absorbing it into fermion operators). However, if one starts with the parent Hamiltonian Eq.~(\ref{total_hamiltonian}) and 
        introduces complex phases to $\Delta$ and $\Delta_v$,  then only one of these two phases can be gauged out (the complex phases of $\Delta$ and $\Delta_v$
    cannot be simultaneously absorbed into fermion operators). Below we will work in the gauge where $\Delta$ is real while $\Delta_v$ is a complex number: $\Delta_v=|\Delta_v|e^{i\theta}=\Delta^R_v+i\Delta^I_v$. The parameter $\theta$ is thus the phase difference between $\Delta$ and $\Delta_v$. The corresponding Hamiltonian  reads 
    \begin{eqnarray}
    \label{model1}
    \nonumber
    H_\text{interchain}&=&\frac{i}{2}\sum_j(\Delta^R_v-t_v)c_{2j-1,1}c_{2j,2}\\
    &+&\frac{i}{2}\sum_j(\Delta^R_v+t_v)c_{2j,1}c_{2j-1,2} \nonumber\\
    &+&\frac{i}{2}\sum_j \Delta^I_v(c_{2j-1,1}c_{2j-1,2}- c_{2j,1}c_{2j,2}) 
    \end{eqnarray}
    In this model, the TRS breaking field $B$ is identified with $B\equiv\Delta^I_v/2t$, as the vertical complex pairing 
        terms, $\propto c_{m,1}c_{m,2}$, do break the TRS.
        
    The Hamiltonian, $H$, is represented in momentum space as follows
    \begin{eqnarray}
    H&=&\frac{1}{2}\sum_k \hat{\psi}^{\dagger}_k h(k) \hat{\psi}_k,\\ \nonumber
    h(k)&=&(-2t \cos k-\mu)\sigma_z\mathop{\otimes} \mathbbm{1}_2+2\Delta \sin k\sigma_y \mathop{\otimes} \mathbbm{1}_2  \\
    &-&t_v\sigma_z \mathop{\otimes} \tau_x+\Delta^R_v \sigma_y \mathop{\otimes} \tau_y -\Delta^I_v \sigma_x \mathop{\otimes} \tau_y.
    \end{eqnarray}
     Complex $\Delta^I_v$ completely changes the structure of the single-particle excitation spectrum. We apply the Bogoliubov transformation to diagonalize $h(k)$ to get the spectrum $E_{s}(k)$ as
    \begin{eqnarray}
    E_{s}(k)&=&\sqrt{\xi^p_1(k)+2s\xi^p_2(k)} \label{pairingenergy}
    \end{eqnarray}
    where $\xi^p_1=(2\Delta \sin k)^2+(2t\cos k+\mu)^2+t^2_v+|\Delta_v|^2$, $\xi^p_2=\sqrt{4\Delta^2 \sin^2 k(\Delta^R_v)^2+|\Delta_v|^2t_v^2+(2t\cos k+\mu)^2t_v^2}$, $s=\pm$.
    As the next step, we proceed with the identification of the 0-MF, 2-MF, and 4-MF phases. This can be achieved by following the procedure outlined for the parent Hamiltonian in
        Sec.~\ref{full}: upon solving the equation $E_{s}(k)=0$, one obtains 
    \begin{eqnarray}
    \nonumber
    (2t\cos k+\mu)^2+|\Delta_v|^2-t^2_v&=&0\\
    (\Delta^I_v\sin k)^2+(2t\cos k+\mu)^2 \sin^2 k&=&0.
    \end{eqnarray}
    Then the phase boundaries of the phase diagram are given in terms of parameters $|\Delta_v|$, $t$, $\mu$, and $t_v$ satisfying the equation 
    \begin{eqnarray}
    |\Delta_v|^2+(2t\pm\mu)^2&=&t^2_v.
    \end{eqnarray}
    Along these boundaries, the spectral gap closes at momenta $k=0,\pi$.
    
    \begin{figure*}
        \centering
        \subfloat[][Complex vertical pairing potential]{\includegraphics[scale=0.37]{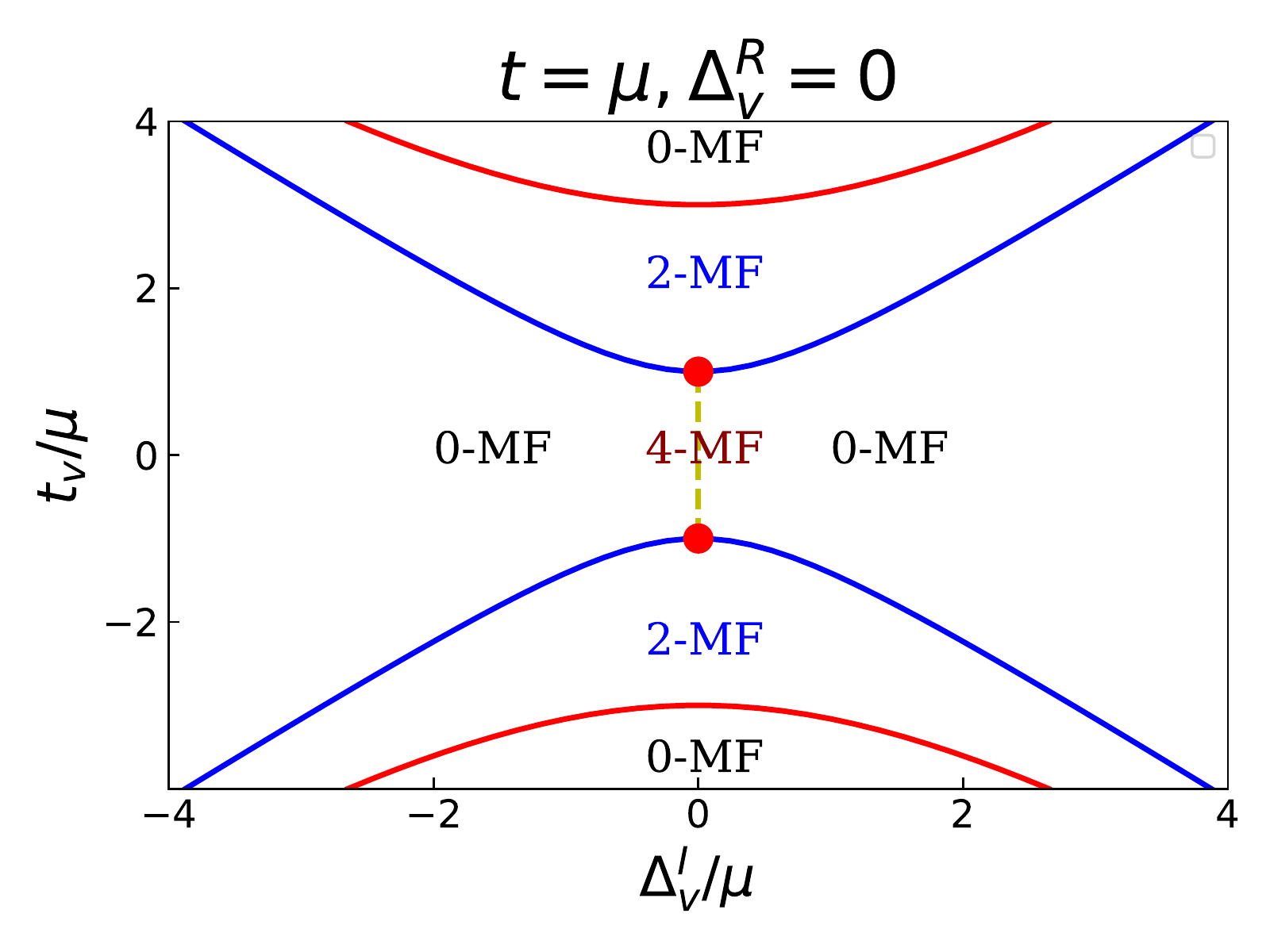}\label{pairingphase}}
        \subfloat[][Uniform Flux]{\includegraphics[scale=0.37]{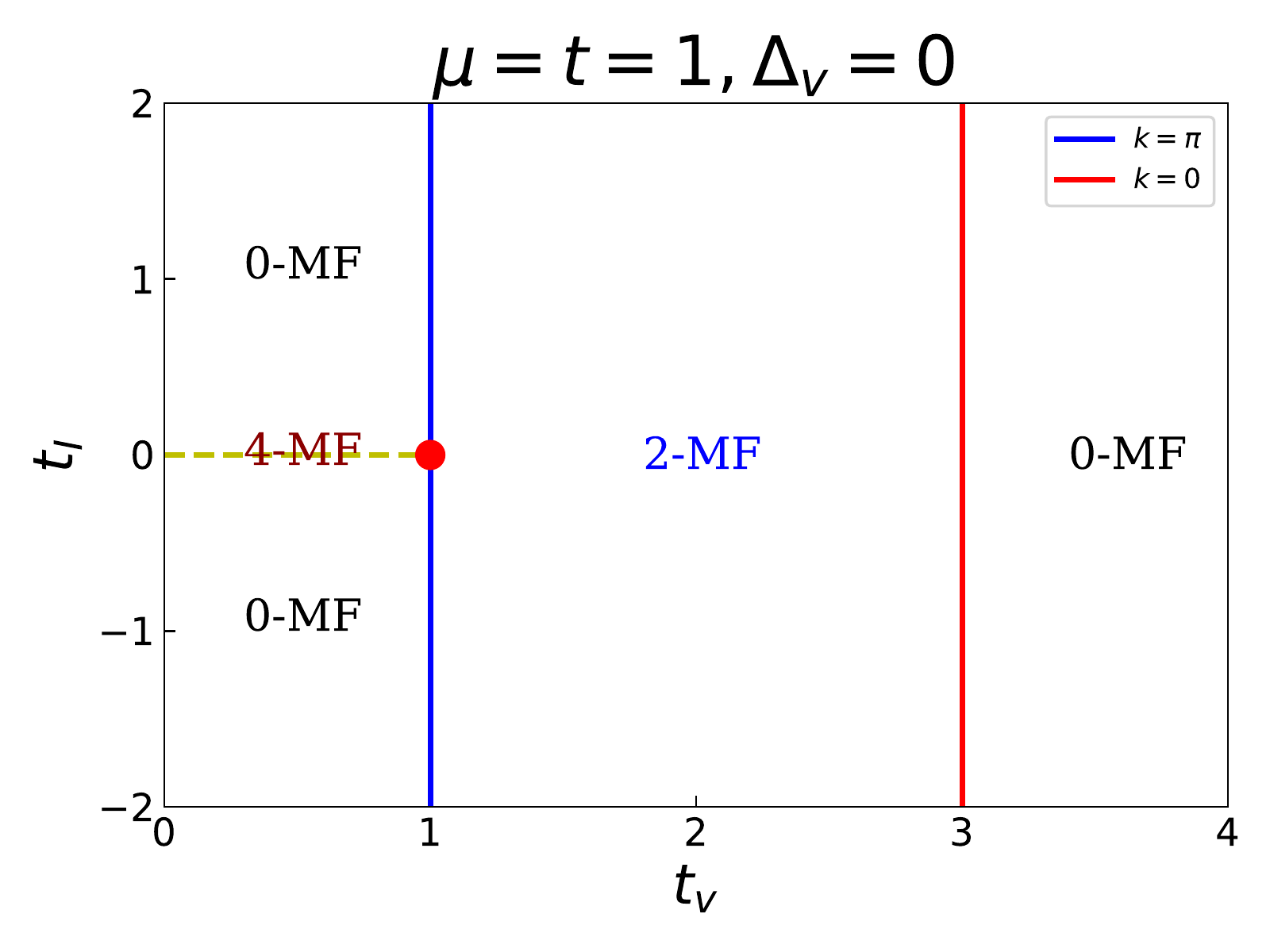}\label{uniformphase}}
        \subfloat[][Staggered Flux]{\includegraphics[scale=0.37]{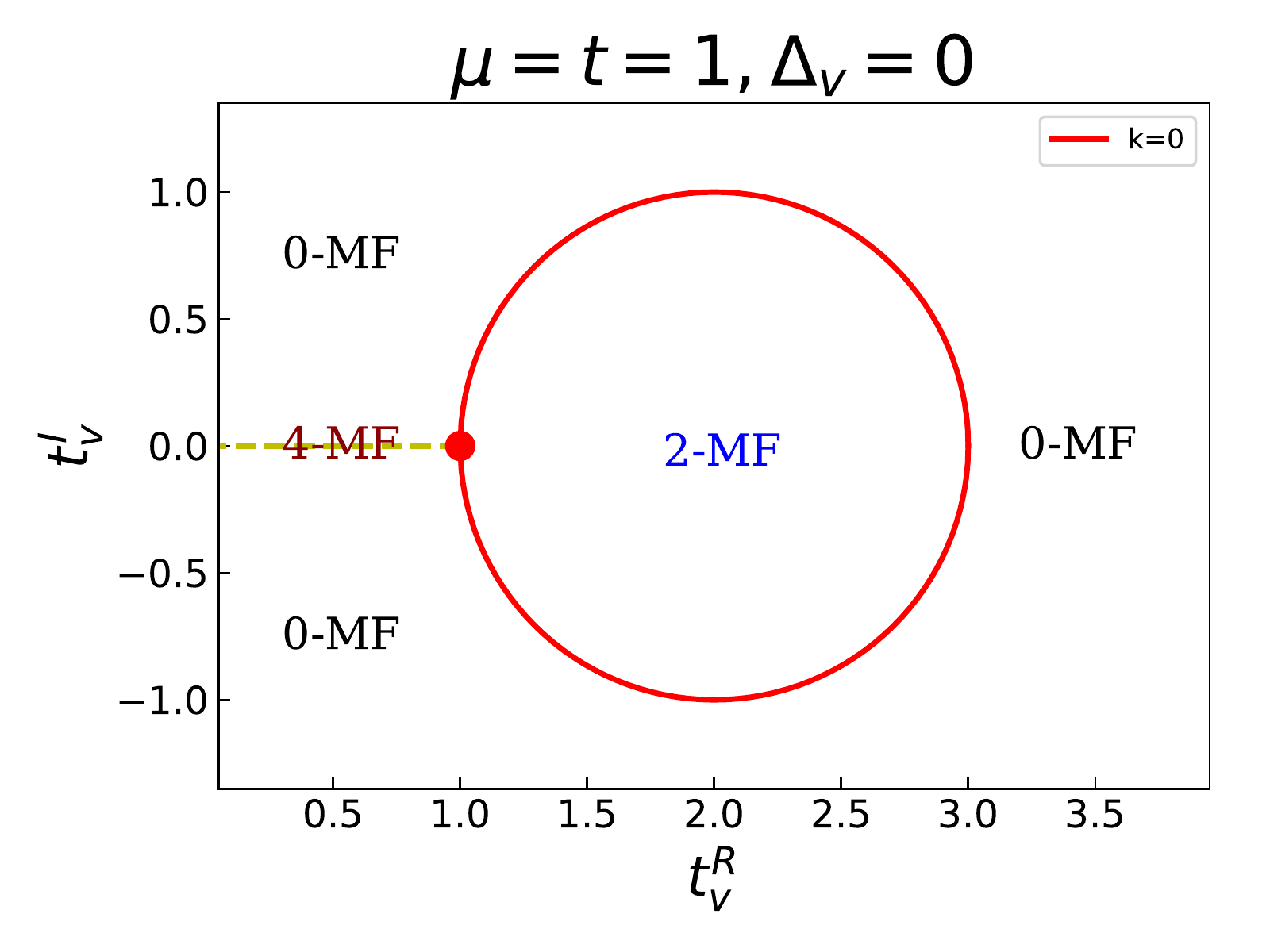}\label{staggerphase}}
        \caption{(Color online) Phase diagrams of (a) Model I, (b) Model II, and (c) Model III. All three phase diagrams support the tricriticality, marked by a bold (red) dot. It separates $4$-MF, $2$-MF and $0$-MF phases. 
        In all three diagrams, the $4$-MF phase resides on the dashed (yellow) line. It is smoothly connected to $0$-MF when the field $B$ is added, while $2$-MF is robust to TRS-breaking perturbation.  }
            \label{PDs}
        \end{figure*}
        
        As the next step, we diagonalize the Hamiltonian (\ref{total_hamiltonian}) with $H_{\text{interchain}}$ given by Eq.~(\ref{model1}) in the Majorana basis with open boundary conditions. Subsequently we numerically analyse the eigenvalues of the Hamiltonian and 
                obtain complete information on the zero-energy boundary modes. The phase diagram is plotted 
                in Fig~(\ref{pairingphase}) in the space of  rescaled parameters $w_v$ and $\Delta^I_v$. 
        All three different phases, namely $4$-MF, $2$-MF and $0$-MF phases exist in this phase diagram: 
        \begin{itemize}
        
        \item The $4$-MF, is an SPT phase characterized by the  $\mathbb{Z}^T_2$ quantum number. It can be smoothly connected to the topologically trivial phase without closing a gap in the presence of the TRS-breaking field. Thus the  $4$-MF SPT phase resides only on the dashed (yellow) line of the diagram, while to the left/right of it, the TRS is broken, and the phase is trivial. 
        \item The $2$-MF phase is topologically ordered \cite{jop15wen}. It is robust to TRS breaking perturbation, and the Majorana modes are immune to $B$.
        
        \item The $0$-MF phase is topologically trivial.
        
        \end{itemize}

        \subsection{Model II: Majorana ladder in a uniform magnetic field}  \label{uniform_flux_sec}
\begin{figure}
            \includegraphics[scale=0.4]{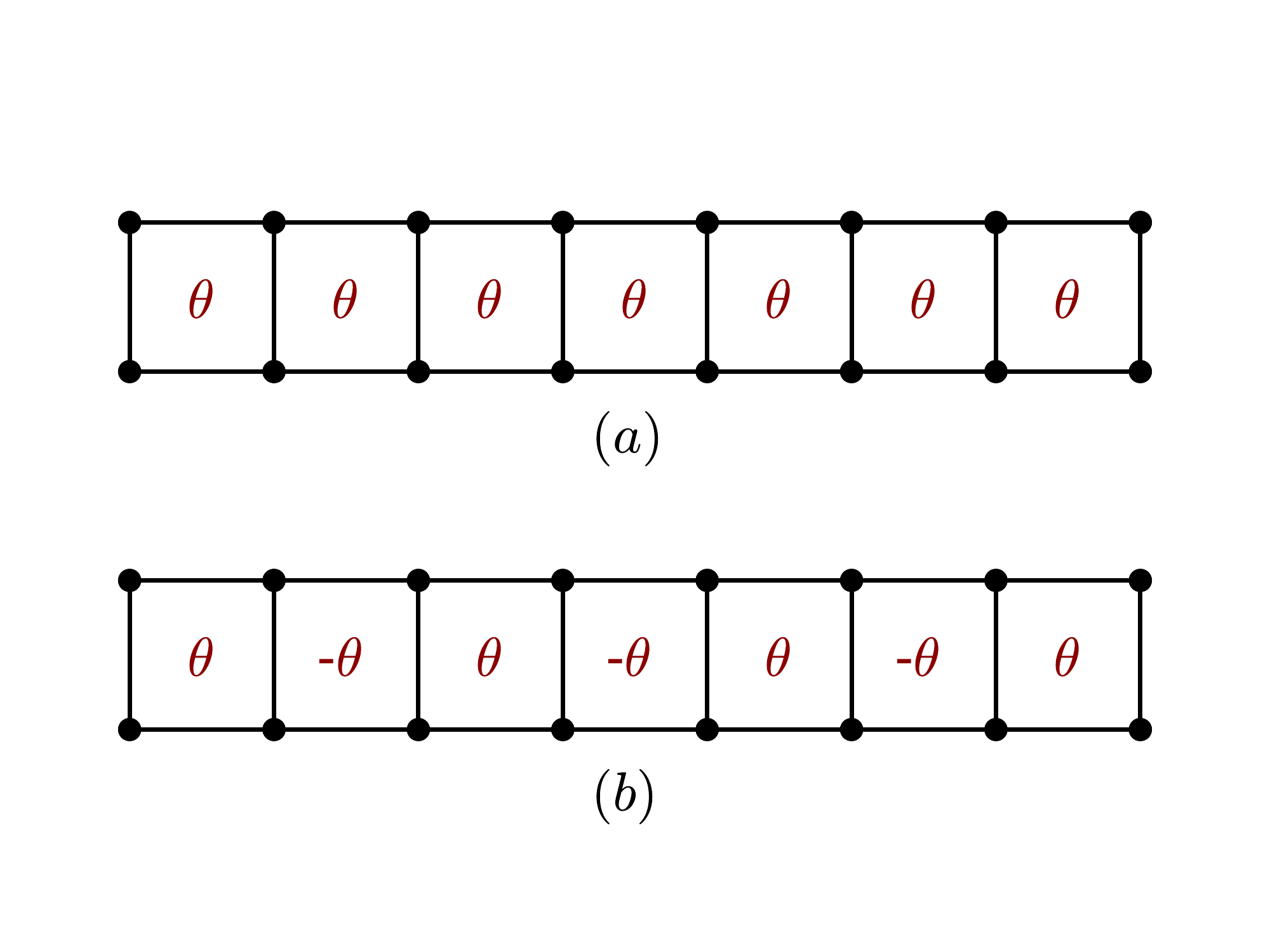}
            \centering
            \caption{(Color online) Parent model in the presence of an external TRS breaking flux field. (a) Uniform magnetic field. 
            (b) Staggered magnetic field.}
            \label{flux}
        \end{figure}

        The ladder model has square plaquettes that can be threaded by magnetic fluxes and around which a fermion can rotate. In this subsection, we will consider a model that corresponds to the parent Hamiltonian (\ref{total_hamiltonian}) in the presence of an external constant magnetic field.   Thus each square plaquette of the ladder is penetrated uniformly by the flux, $\theta$, as shown in Fig. (\ref{flux})a.
        The gauge field, which generates the uniform flux, couples to fermion hoppings along the links. For simplicity, we chose to work in the gauge where the complex phase is added to the intra-chain hopping, $t$:  $te^{i\theta/2}$ in $H_1$ and $te^{-i\theta/2}$  in $H_2$. The Hamiltonian of Model II is thus given by Eq.~(\ref{total_hamiltonian}), where $H_{\text{iterchain}}$ is unchanged and 
        \begin{eqnarray}
        \nonumber
        H_{1}=\sum_{j} &\mu& \hat{a}^{\dagger}_{j,m}\hat{a}_{j,m}+(\Delta \hat{a}_{j,m}\hat{a}_{j+1,m}+h.c.)\\
        &-&( t e^{i\theta/2} \hat{a}^{\dagger}_{j+1,m}\hat{a}_{j,m}+h.c.)\\
        \nonumber
        H_{2}=\sum_{j} &\mu& \hat{a}^{\dagger}_{j,m}\hat{a}_{j,m}+(\Delta \hat{a}_{j,m}\hat{a}_{j+1,m}+h.c.)\\
        &-&( t e^{-i\theta/2} \hat{a}^{\dagger}_{j+1,m}\hat{a}_{j,m}+h.c.).
        \end{eqnarray}
        Here the intra-chain terms $H_{1,2}$ reduce to the parent Eq.(\ref{parent_intra}) at $\theta=0$. We remind the reader that the uniform flux $\theta$ breaks the \text{TRS} and thus the 
        TRS breaking field $B$ here is identified with $B\equiv\frac{1}{2} \sin (\theta/2)$.
        The momentum space representation of the Hamiltonian in the Majorana basis helps to trace the modification of i) the boundary modes, and ii) of the bulk spectrum. The technical details of analytical calculations are presented in Appendix(\ref{uniform_app}).
        
        The phase diagram of the uniform-flux model under consideration contains three different phases: a trivial 0-MF phase, a topologically ordered 2-MS phase, and an SPT 4-MF phase that is situated on a line. All three phases come together, giving rise to a tricritical point. 
        The two phase boundaries of the model are given by the equation 
        \begin{eqnarray}
        (2t_R\pm\mu)^2&=&t^2_v-\Delta_v^2
        \end{eqnarray}
        where $t_R=t \cos \theta/2$. It is clear that $t_I$ does not influence the phase boundary equation in any way. The phase diagram in the space of parameters $t_I$ and $t_v$ is shown in Fig.~\ref{uniformphase}.  
        
        \subsection{Model III: Majorana ladder in a staggered magnetic field}  \label{stagger_flux_sec}
        One can break the TRS also by introducing a staggered magnetic field, instead of the uniform magnetic flux discussed in the previous section. Here we consider a staggered $\theta$ flux threading each of the square plaquettes of the ladder, whose signs are alternating. An example of a certain lattice segment with fluxes is shown in Fig.~\ref{flux}b. The staggered disposition of the sign of $\theta$ implies doubling of the unit cell, the net flux through which is zero. 
        The corresponding single particle Hamiltonian, based on the parent model \ref{total_hamiltonian}, can be written in gauge where alternating complex phases are attached to vertical interchain hopping parameters: $t_v\rightarrow t_v e^{\pm i\theta/2}$ with alternating signs. Thus, the full Hamiltonian is still given by  Eq.~(\ref{total_hamiltonian}), with the interchain Hamiltonian being
        \begin{eqnarray}
        \nonumber
        \label{h3}
        H_{\text{interchain}}=&-&\sum_{j=2n-1} t_v e^{i\theta/2} \hat{a}^{\dagger}_{j,1}\hat{a}_{j,2}-\sum_{j=2n} t_v e^{-i\theta/2} \hat{a}^{\dagger}_{j,1}\hat{a}_{j,2}\\
        &+&\sum_{j} \Delta_v \hat{a}_{j,1}\hat{a}_{j,1}+ h.c., 
        \end{eqnarray} 
        where the staggered field $\theta$ breaks the TRS.  Thus, we identify the TRS breaking field $B$ here with $B\equiv t_v\sin (\theta/2)/(2t)$.
        
        As the next step, we study bulk and boundary spectra of the model. To investigate the boundary modes, we switch to the Majorana basis and find the spectrum using the method outlined in Appendix A. 
        To derive the bulk spectrum, we perform Fourier transformation in Eq.~(\ref{h3}), which diagonalizes the Hamiltonian. We present the corresponding calculation in Appendix (\ref{stagger app}).
        
        The outlined analysis helps us to study the phase diagram of the model systematically. The phase boundaries separating the trivial 0-MF phase, a topologically ordered 2-MF phase, and an SPT 4-MF phase are given by 
        \begin{eqnarray} 
        (2t\pm t^R_v\pm\Delta_v)^2+(t^I_v )^2=\mu^2.
        \end{eqnarray}
        Here $t^I_v$ does enter into the expression for phase boundaries, which now in the space of parameters $t^I_v$ and $t^R_v$ becomes very interesting. It is shown in Fig.~(\ref{staggerphase}), from which we see that it is topologically different from the one corresponding to the uniform magnetic field. In particular, the space corresponding to the 2-MF state is now compact. 
        
        \subsection{Phase diagrams of models with broken TRS: the tricriticality separating 4-MF, 2-MF and 0-MF phases}
        
        In this subsection, we will discuss common features of all three phase diagrams corresponding to the models I, II, and III defined above. In each phase diagram, shown in Fig.~(\ref{pairingphase}), Fig.~(\ref{uniformphase}), and Fig.(\ref{staggerphase}), there is a peculiar $4$-MF phase, which resides on a dashed (yellow) line. The dashed line itself also represents a first-order phase transition line. It has the following description: when the TRS breaking field, $B$, is added to the $4$-MF state, the ground state degeneracy is \textit{lifted}. This happens because the zero-energy boundary modes obtain finite energy proportional to $B$. Then the spectrum develops a level-crossing, signaling the first-order phase transition. 
        There are phase boundaries corresponding to the second-order phase transitions between $2$-MF and trivial phases shown in Fig.~(\ref{PDs}) by full lines (blue and red).
        Interestingly enough, the tricriticality happens as the intersection of the first-order transition line and the second-order transition line.
         The study of the {\it universal} properties around this tricriticality, shown by bold (red) dots in Fig.~(\ref{PDs}), is the main focus of the present paper. 
        
        The tricriticality separates three different phases: $4$-MF(characterized by $\mathbb{Z}^T_2$ topological invariants of the SPT theories), $2$-MF(fermionic topologically ordered phase) and $0$-MF (trivial phase).
        
        \begin{itemize}
            \item         In $4$-MF phase, near the tricriticality, two Majorana edge modes have localization lengths $\xi_+$ while the other two have localization length $\xi_-$, see Eq.(\ref{correlation_length}) for definitions. 
        
        \item  The Majorana edge modes in a $2$-MF phase, near the tricriticality, are characterized only by one localization length. 
        
        \item At the tricriticality, there are two Majorana edge modes, with the same localization lengths $\xi$. As one departs from tricriticality in Fig.~(\ref{PDs}), and moves along the gapless phase boundary (the blue line), one immediately enters the trivial gapless phase with no boundary Majorana modes. To achieve such a nontrivial transition, one will have to tune the model parameters correspondingly.
        Unlike this critical line, the tricriticality does support two Majorana modes, and as such, it represents an example of the gapless SPT phase.

                \end{itemize}
                
                The phase diagram Fig.(\ref{pairingphase}) has two tricritical points positioned at 
        $(\Delta^I_v, t_v)=(0,1)$ and $(\Delta^I_v, t_v)=(0,-1)$ (measured in units of $\mu=t$). The Majorana localization lengths at the former is $\xi_{-}$ while at the latter it is $\xi_{+}$. 
        In two other TRS breaking situations,  only one tricritical point is shown, however, there are two such points as the phase diagrams are symmetric with respect to the vertical axis. 
        
        In all three situations with TRS breaking and the tricriticality, we define the universal dimensionless scale $g$ as:
        \begin{eqnarray}
        g=\alpha(\xi)\sqrt{N}B  \label{scale_early}.
        \end{eqnarray} 
        
        The function, $\alpha(\xi)$, is model dependent. 
        The reason it enters into the universal scale, $g$, is the following. Besides the diverging correlation length, $\xi_{\textbf{cor}}=\infty$, there is one more length scale, the localization length $\xi$, around the tricriticality. The models under consideration are described by short-ranged hoppings and pairings. These short-ranged terms are characterized by the lattice constant $a$, which is set to be unity throughout this paper. The universal phenomena at criticalities emerge when $1 \ll \xi_{\textbf{cor}}$. However, the new length scale $\xi$, which appears to be finite, needs to be carefully treated and compared with the existing scales: i) when $\xi \ll 1\ll \xi_{\textbf{cor}}$ the short-ranged properties, characterized by the lattice spacing ($\sim 1$), can be captured by neither the localization length $\xi$ nor the correlation length $\xi_{\textbf{cor}}$. Thus  $\alpha(\xi=0)\equiv 1$, which is universal for all three models;   ii) when $\xi \sim 1$, i.e, the localization length is comparable with lattice spacing, the short-ranged physics matters. That is the reason why the model-dependent function $\alpha(\xi)$ determines the scale, $g$.

        In the next section, we will show how the scale $g$ naturally emerges in the universal finite-size scaling corrections to ground state energy.

	    \section{Universal Finite size scaling effect around the tricriticality} \label{finite}

        In this section, we will show the emergence of the universal scale $g$ describing the finite size correction to the ground state energy around the tricriticality. Namely, we will show that the finite size correction to the ground state energies of models I, II, and III has the universal form given by Eq.~(\ref{Ke_scaling}).
        
        Further, we will show that in two simultaneous  double-scaling\cite{prb76Wu} limits: a) $N\rightarrow\infty$, $m\rightarrow0$ with $w=const$ and b) $N\rightarrow\infty$, $B\rightarrow0$, with $g=const$, the function $f(w,g)$ is a universal function of two variables
        (in Eq.~(\ref{Ke_scaling}). To uniquely identify $f(w,g)$ around the tricriticality, we adopt the following convention: we chose the sign of $m$ to be $m>0$ for $4$-MF and for $0$-MF, while $m<0$ for $2$-MF).
    
        \subsection{Numerical approach to compute the finite-size scaling function} \label{method_f}
    Here we outline the numerical approach for the calculation of the ground state energy, $E(N,m, B)$. We obtain it by performing a summation of the occupied single-particle energy levels corresponding to the Hamiltonian under open boundary conditions. The per-particle average bulk energy, $\epsilon(m, B)$, is obtained upon summing up the occupied energies and dividing the result by $N$. This procedure yields
        \begin{eqnarray}
        \epsilon&=&-\frac{1}{4\pi}\int_{BZ} dk   \sum_{s} E_s(k)
    \end{eqnarray}
    where BZ stands for the  Brillouin zone,  $E_s(k)$ is single-particle excitation energy in the $k$-space corresponding to band $s$. For example, in case of model I, one has 2 energy bands and thus $s$ acquires only two values $s=\pm$ (for this model, $E_s(k)$ is given by Eq.(\ref{pairingenergy})). The boundary energy, $b(m,B)$, is then given by
    \begin{eqnarray}
    b&=&\lim_{N\rightarrow \infty} \left( E(N,m,B)-N\epsilon(m,B) \right).
    \end{eqnarray}
Having identified all the necessary ingredients, the finite size scaling function $f$ is obtained from
    \begin{eqnarray}
    f&=& \lim_{n\rightarrow \infty}  n\cdot \left( E(n,m,B)-n\epsilon(m,B)-b\right) \label{exact_f}.
    \end{eqnarray}
    Numerically, we pick a large system with $N_0=1000$ and compute
        $\tilde{b}=E(N_0,m,B)-N_0\epsilon(m,B)$.
  Having evaluated the boundary energy, $\tilde{b}$, we take $n_0=100$ to evaluate the finite size scaling function, labeled by $\bar{f}$: $\bar{f}= n_0\cdot \left( E(n_0,m,B)-n_0\epsilon(m,B)-\tilde{b} \right)$.
    The outlined computation is based on two  approximations: i)The evaluated boundary energy $\tilde{b}$ generates an error $e_1\sim\mathcal{O}(n/N)$ that contributes to $\bar{f}$. ii) The higher order terms are ignored, which yields an error $e_2~\mathcal{O}(n^{-1})$ to $\bar{f}$. The total error is thus $f-\bar{f}\sim \mathcal{O}(n/N)+\mathcal{O}(n^{-1})$. By keeping track of these errors and accumulating statistics, one can obtain an
    accurate data for $f$ with controlled precision.

\subsection{Emergence of the new scale $g$}
\begin{figure*}
    \centering
    \subfloat[][]{\includegraphics[scale=0.45]{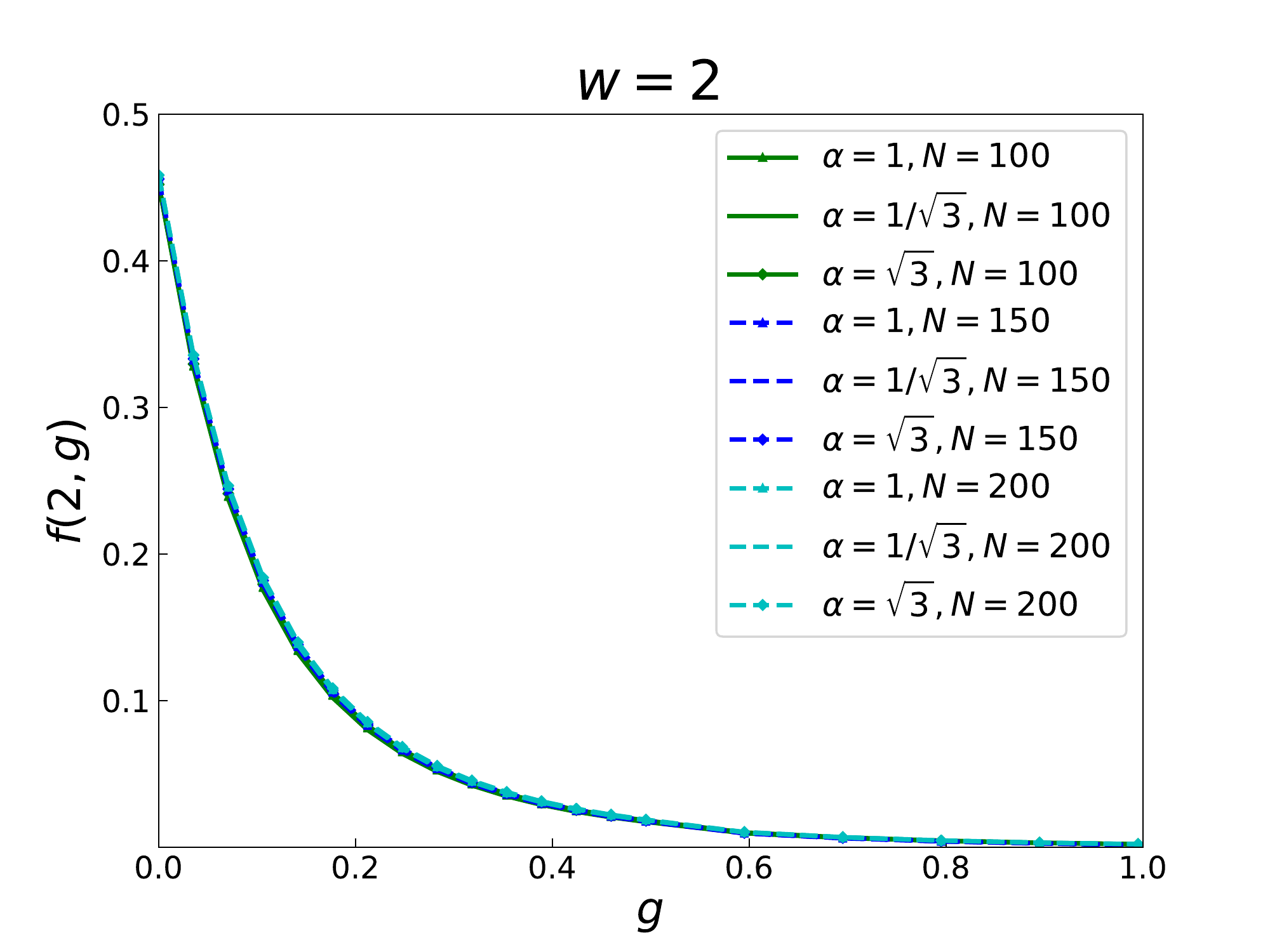}\label{w2}}
    \subfloat[][]{\includegraphics[scale=0.45]{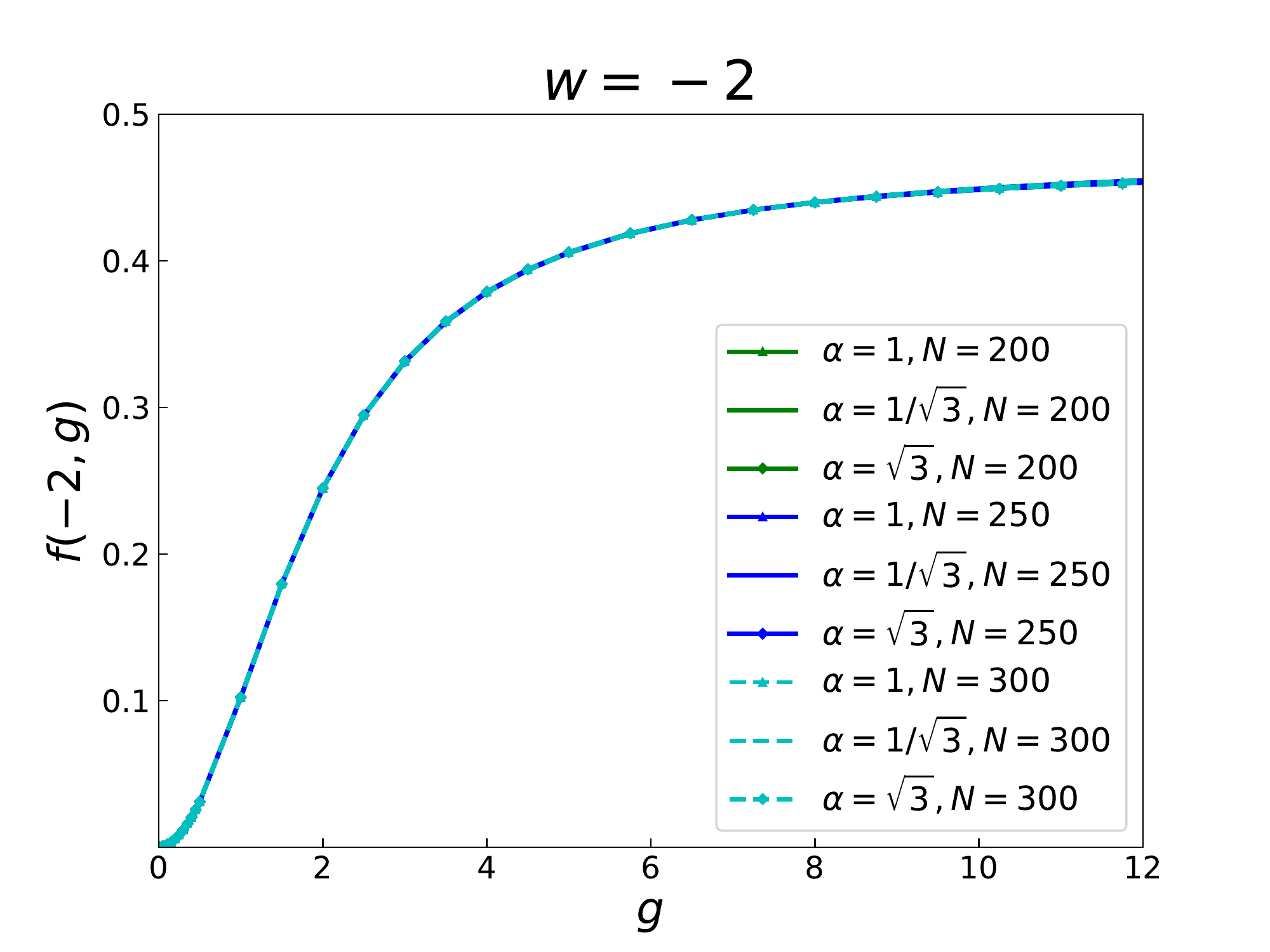}\label{n2}}
    \caption{(Color online) The finite size scaling function $f$ is plotted as a function of $g$ for Model I. Fig.(\ref{w2}) corresponds to $w=2$ and Fig.(\ref{n2}) corresponds to $w=-2$.
    These curves have been obtained for  nine different values of $\alpha=\sqrt{\coth1/2\xi}$ and $N$. 
    } 
    
    \label{steady_state}
\end{figure*}
In this subsection, we take the model I: coupled Majorana chains with {\it complex vertical pairing potential} as an example and use the method of Sec.(\ref{method_f}) to evaluate its finite scaling $f$. Then we show that $f$ is a double scaling function of two variables, the standard variable $Nm$ and a completely new emergent scale $g$.
    
There are several important quantities around the tricriticality: i)The spectral gap $m$, (which is identified with the mass of the low-energy effective field theory); ii) The localization length, $\xi$, of Majorana edge modes existing at the tricriticality, which characterize the topological properties of the tricritical point; iii) The system size, $N$, which is the essential ingredient for studies of finite size effects; iv) The symmetry-breaking field $B$, which induces the tricriticality.
We find that the finite scaling $f$ does explicitly depend on certain combinations of $m$, $N$, $\xi$ and $B$: if we vary $m$, $N$, $\xi$ and $B$
keeping $g=\sqrt{N\coth (1/2\xi)}B$ (for Model.I) and $w=Nm$ constant, $f$ always yields exactly the same value.  This means that $f\equiv f(w,g)$ is only a function two variables, $w$ and $g$. 
This observation puts forward the definition of the scale $g$ in Eq.(\ref{scale_early}). The scale $w=Nm$ is consistent with previous papers on the finite-size scaling effect with no TRS breaking \cite{prl86Cardy,prl16Gulden}. 

Figs.(\ref{w2}) and (\ref{n2})  give concrete results to support that $f$ is a function of $g$: $f$ yields the same value if we keep $g$ to be constant simultaneously changing $N$, $\xi$, and $B$. 
In the remainder of the paper, we will use $f(w,g)$ to represent the finite-size scaling around the tricriticality.

\subsection{Universality of the finite size scaling function $f(w,g)$}
We calculate the scaling function $f$ for all three models introduced in Sect.(\ref{breaking}). We show the universality nature of $f(w,g)$ for all of them. In each model, symmetry-breaking $B$ is realized in different ways by involving different  lattice parameters: $\Delta^I_v$ is imaginary part of the vertical pairing potential for model I in Sec.(\ref{comlex__pairing_potential_section}), $t^I$ is imaginary part of the intra-chain hopping for model II in Sec.(\ref{uniform_flux_sec}), and $t^I_v$ is imaginary part of vertical hopping for model III in Sec.(\ref{stagger_flux_sec}). 

We find that all three models yield the same finite-size scaling function $f(w,g)$, although the field $B$ has different origins in them. Fig.(\ref{central}) depicts $f(w=0,g)$ at criticality for all three models, and Fig.(\ref{uw2}), depicts $f(w=2,g)$ corresponding to the gapped phase for all of them. These plots support the universality, either at the criticality or for the gapped phase. The finite size scalings $f(w,g)$, calculated for three models, provide the same sets of curves.

\subsection{Features of the finite size scaling function $f(w,g)$}  
Here we analyse the features of $f(w,g)$. The function itself is given by a 2D surface in a 3D space of $(w,g,f)$.  We depict some of the curves residing on this surface: for each value of $w$, we plot $f(w,g)$ as function of $g$. We use the parameters of the complex-vertical pairing model I as an example to present $f(w,g)$ (although $f$ is universal, for simplicity of the presentation we work in the parameter space of this model).  

\begin{figure}
            \includegraphics[scale=0.5]{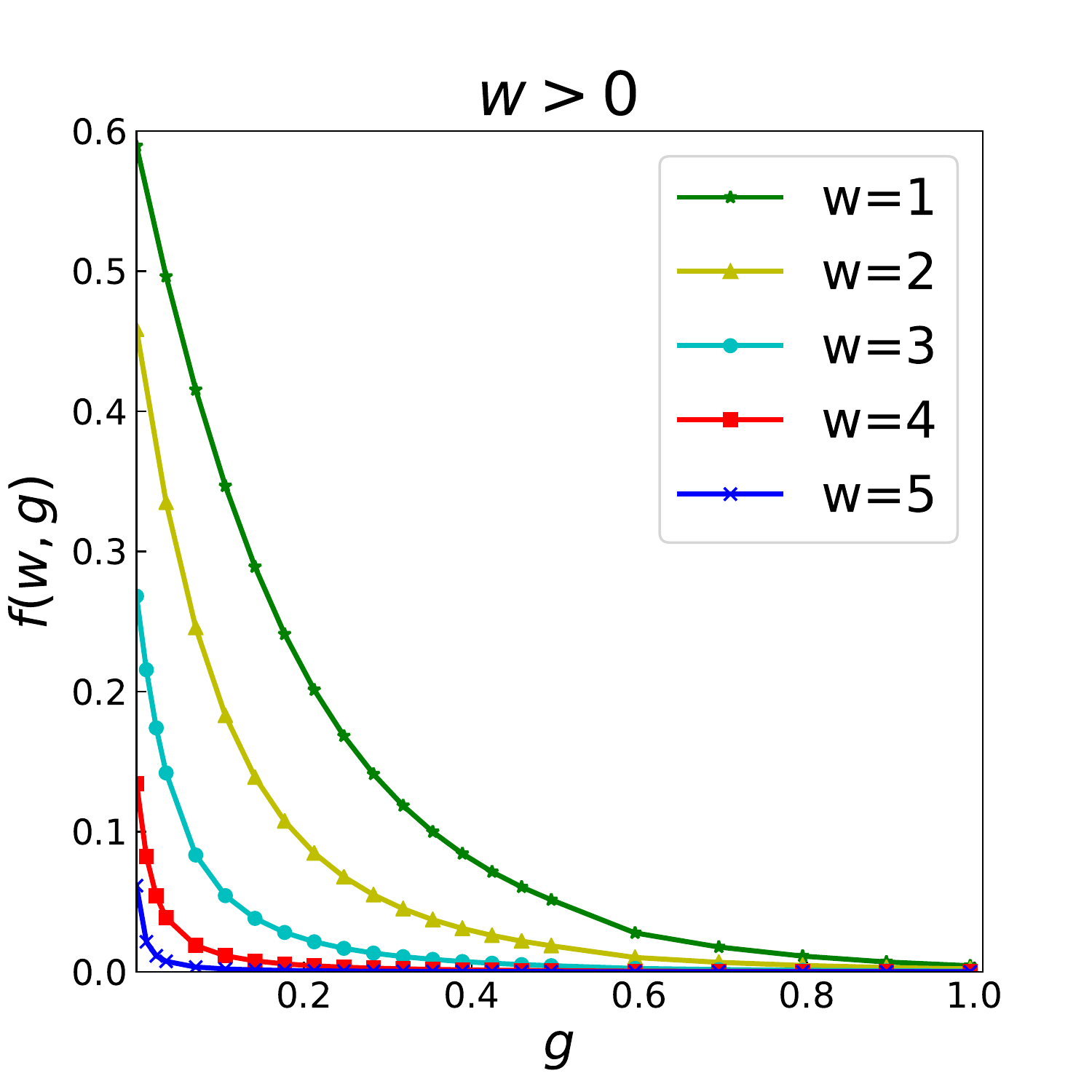}
            \centering
            \caption{(Color online) The finite size scaling functions at $w\geq1$.  Curves from top to bottom correspond to values $w=1,2,3,4,5$.
            Here only the $g\leq 1$ is presented. As $w$ increases, $f(w,g)$ decays faster since larger $w$ means being further away from second order phase transition. Then, the contribution coming from the first order transition dominates.  
            }
            \label{largep}
        \end{figure}

\begin{figure}
            \includegraphics[scale=0.45]{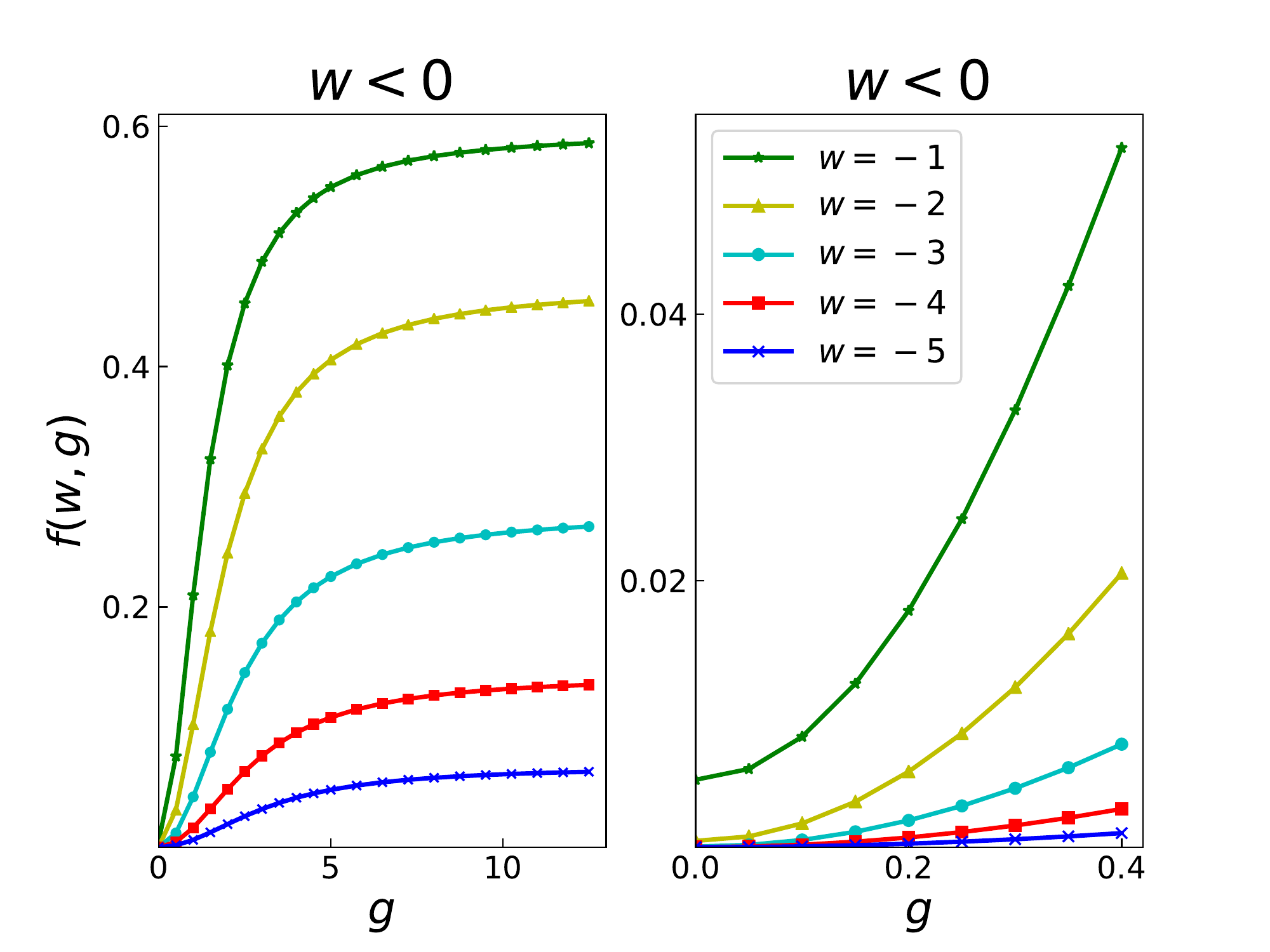}
            \centering
            \caption{\text{(Color online) The finite size scaling function at $w\leq -1$}.  The left figure depicts the scaling function when $g$ varies from $0$ to $12$, while the right figure depicts the behavior when $0\leq g\leq 0.4$. 
            }
            \label{largen}
        \end{figure}

\begin{figure}
            \includegraphics[scale=0.55]{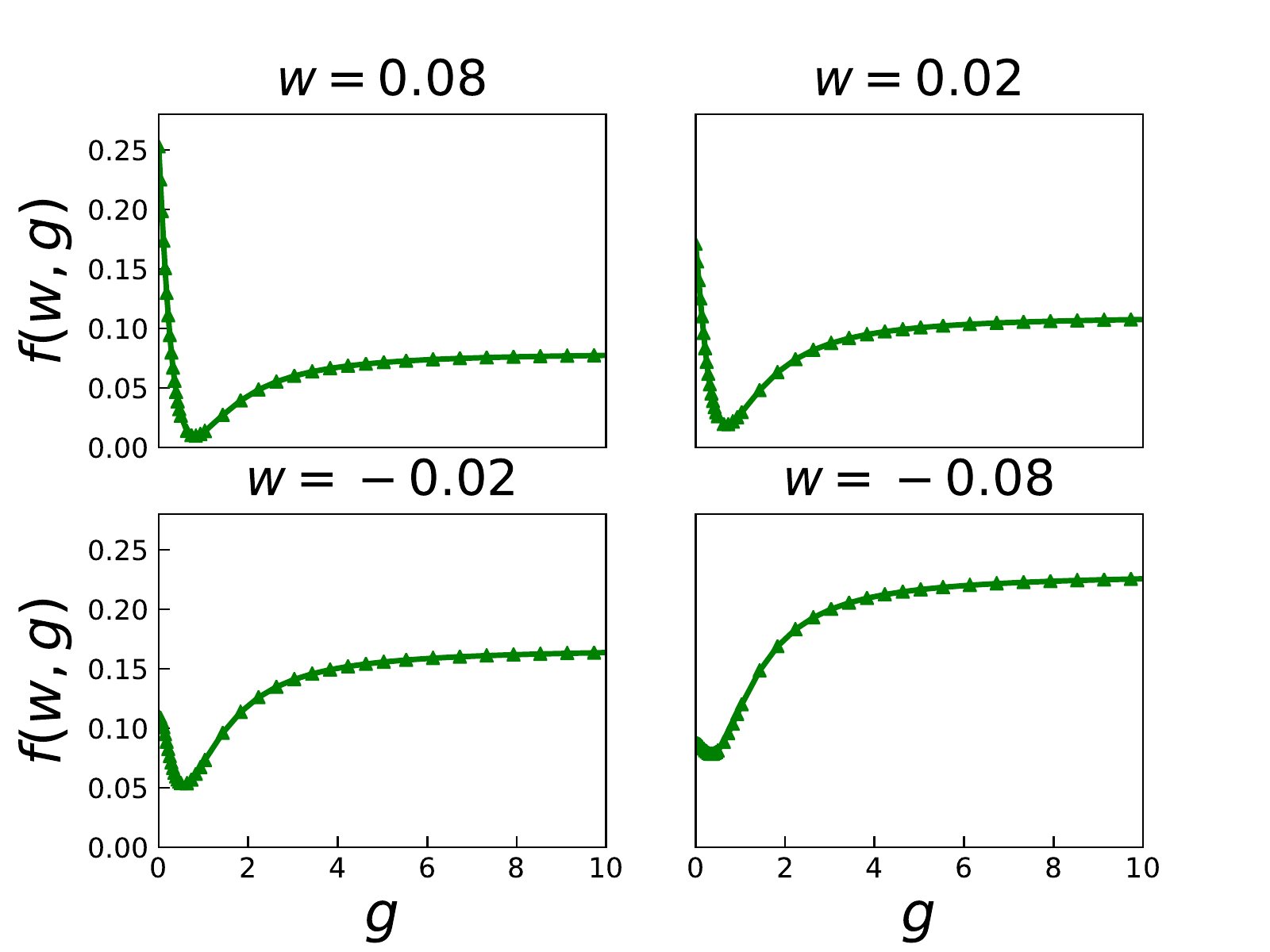}
            \centering
            \caption{(Color online) The finite size scaling function at $-0.1<w<0.1$. 
            }
            \label{sink}
        \end{figure}

\begin{figure}
            \includegraphics[scale=0.42]{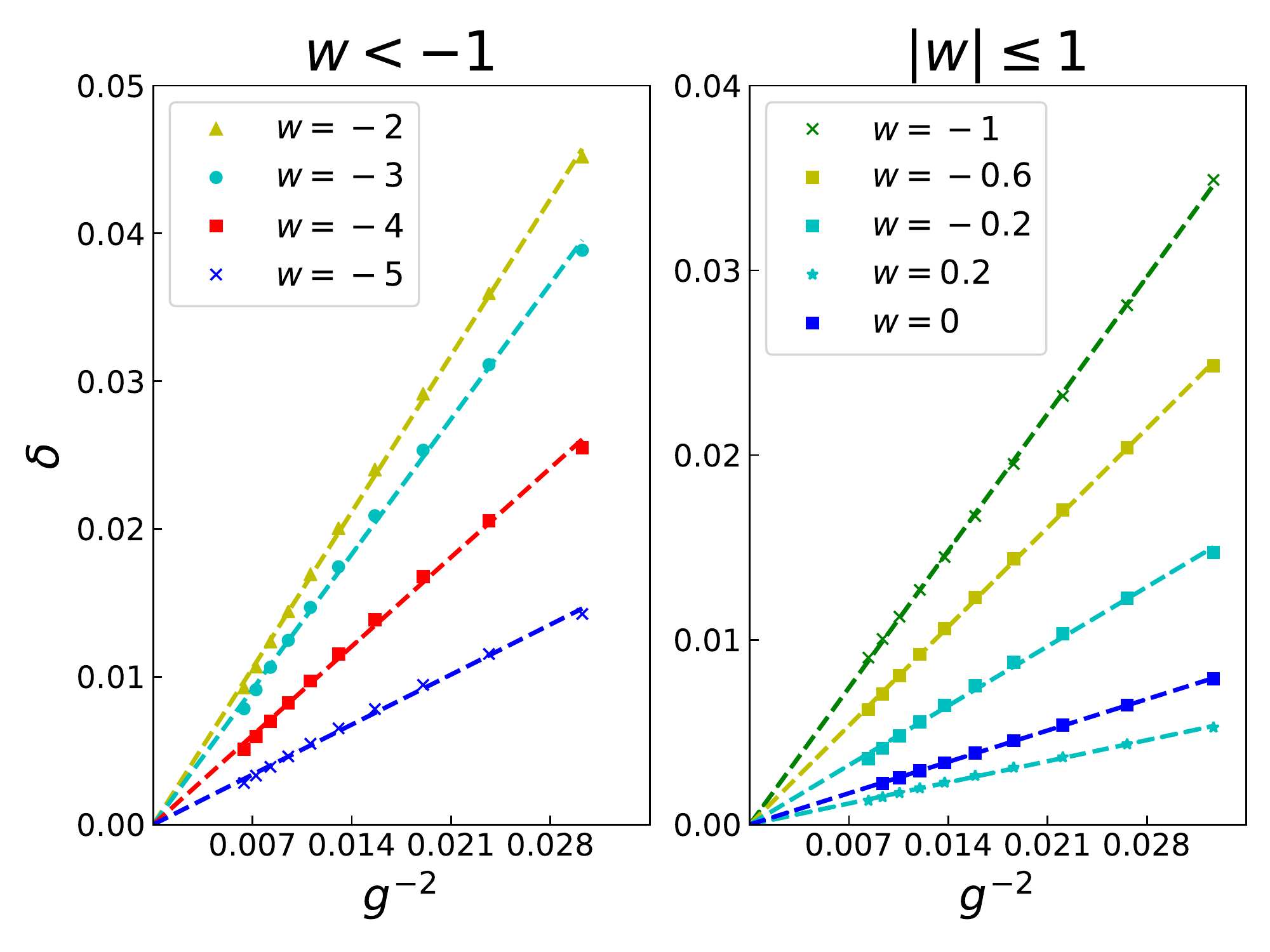}
            \centering
            \caption{(Color online) The linear fit of $f(w,g)$ versus $g^{-2}$ is shown at $g\gg 1$: $f(w,g)=f(w,\infty)-\beta/g^2$. Figures depict the function  $\delta=f(w,\infty)-f(w,g)$ versus $g^{-2}$. The $g$-independent shift $f(w,\infty)$ is given by the constant part in the linear fit.
            }
            \label{linear}
        \end{figure}

\begin{figure}
            \includegraphics[scale=0.5]{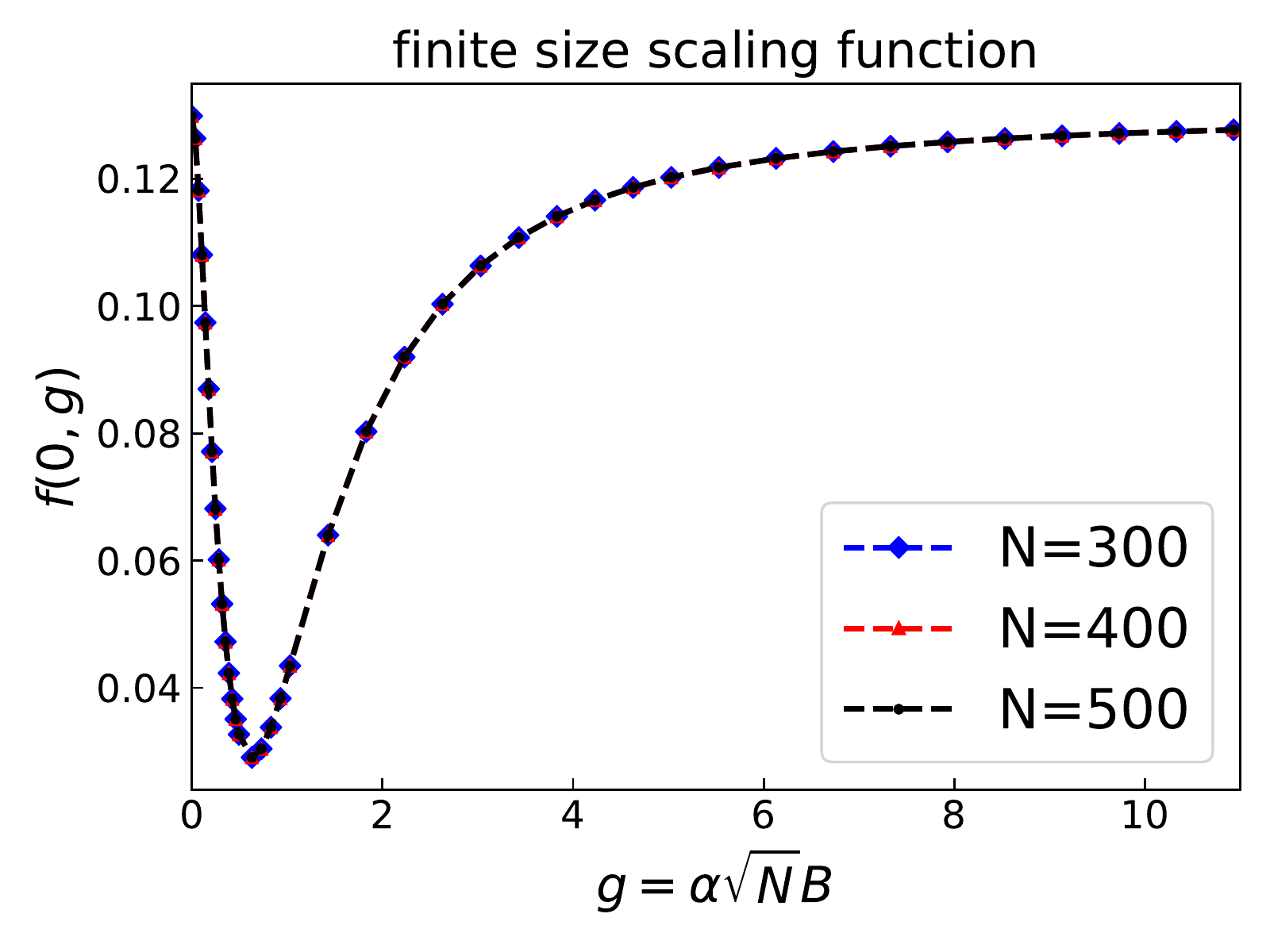}
            \centering
            \caption{(Color online) The finite size scaling function at criticality $(w=0)$ plotted for three different values of $N$.
            }
            \label{centralmore}
        \end{figure}

 The results are as follows: Figs.(\ref{largep}) and (\ref{largen}) show the behavior of $f(w,g)$ with $w$ ranging between $1\leq w\leq 5$ and $-5\leq w\leq -1$. When $w$ belongs to these intervals, $f(w,g)$ is a monotnoic function of $g$.  In the first interval, $f(+|w|,g)$ decays exponentially as a function of $g$, while in the second one, $f(-|w|,g)$ converges slowly (which is in fact $\propto g^{-2}$ at large $g$). Figs. (\ref{sink}) and (\ref{centralmore}) show the behavior of $f(w,g)$ in the interval $-0.1<w<0.1$ For small $|w|<0.1$, $f(w,g)$ is a non-monotnoic function of $g$. At the criticality, $w=0$, the finite size scaling $f(0,g)$ strongly depends on the scale $g$. Fig.(\ref{linear}) shows that the function $f(w,g)$ saturates to a constant value  as $\propto g^{-2}$ at large $g$.

 At this stage, it is interesting to compare our universal finite-size scaling function $f(w,g)$ with the result of Ref.~\cite{prl16Gulden}
for the scaling function, $f_\text{D}(w)$. The latter was calculated in a situation, where there were no tricriticalities.
From all the curves above, we confirm the role of $g$: $g$ provides interpolation between $f_\text{D}(w)$ and $f_\text{D}(-w)$ by $f(w,g=0)=f_\text{D}(w)$ and $f(w,g=+\infty)=f_\text{D}(-w)$.  For example, it is shown in Ref.~\cite{prl16Gulden} that $f_\text{D}(1)\approx 0.6$ in a 1D topological phase while $f_\text{D}(-1)\approx 0$ in a trival phase. The curve with $w=1$ in Fig.(\ref{largep}) provides the decaying function $f(1,g)$ from $0.6$ (that is the value of $f_\text{D}(1))$ to $0$ (value of $f_\text{D}(-1))$.
        
        \section{Low-energy effective theory around the tricriticality} \label{lowformt}
        
    In the section, we start from the parent Hamiltonian $h(-i \partial_x)$, given by Eq. (\ref{BdG}), which supports the low-energy Majorana field theory and two Majorana edge modes. As the first step, we rewrite the symmetry-breaking Hamiltonian $H_{\text{interchain}}(B)-H_{\text{interchain}}(B=0)$, where $H_{\text{interchain}}(B)$ is given by Eq.~(\ref{model1}), in the momentum representation. This yields $h_{B}=-B \sigma_x \otimes \tau_y$, that acts as a perturbation to the parent Hamiltonian. Then we consider the low-energy sector of the perturbed theory by calculating the matrix elements of $h_B$ using the low-energy eigenstates of the full parent Hamiltonian (note that in this basis the parent $h(-i \partial_x)$ is diagonal). 
    At this level, we explicitly show the emergence of the new scale, $g$, in the low-energy effective matrix. Finally, we obtain the spectrum of the effective matrix and analyze the essential properties around the tricriticality.
        
        \subsection{Low-energy sector around the tricriticality at $B=0$} \label{sector}
        
        Here we identify the low-energy eigenstates of the parent Hamiltonian $h(-i \partial_x)$, around the tricriticality. 
        Diagonalization of $h(-i \partial_x)$ yields two energy bands that are above the chemical potential. The lower one corresponds to the low energy states being described by the $1+1$D Majorana field theory. The higher energy band also brings upon states supporting and protecting Majorana zero-energy edge modes. In this way, we obtain two sets of low-energy states: i) states that are described by $1+1$D Majorana field theory. We call these states MFT-states. ii) zero-energy edge modes that are protected by the topological number of the higher energy band of the ladder model. 
        
        We take Model I with the \textit{complex-vertical pairing potential} as an example, and start from the Hamiltonian $h(-i\partial_x)$:
    \begin{eqnarray}
    \nonumber
    h(-i\partial_x)&=&(-2t \cos (-i\partial_x)-\mu)\sigma_z\mathop{\otimes} \mathbbm{1}_2\\
    &+&2\Delta \sin (-i\partial_x)\sigma_y \mathop{\otimes} \mathbbm{1}_2  
    -t_v\sigma_z \mathop{\otimes} \tau_x  \label{low_energy_ladder}
    \end{eqnarray}
    where we took $\Delta_v=0$ for simplicity (as it does not affect the universal properties under consideration). 

    Around the criticality, the wavefunction of the lower band can be obtained from linearization of $h(-i\partial_x)$. The operator $h(-i\partial_x)$ has two energy bands above the chemical potential, with the lowest one being $E_k=\sqrt{m^2+k^2}$ with $m=\left(2t-\mu-w_v \right)/(2t)$, at $\mu,w>0$. The corresponding eigenstates are the MFT-states. 
        For an infinite system (no imposed boundary conditions), these MFT states are characterized by good quantum numbers $k$(momentum) and $n$(energy):
        \begin{eqnarray}
    h(-i\partial_x)e^{ikx} u(n,k)=n\cdot2t\sqrt{m^2+k^2}e^{ikx}u(n,k)
        \end{eqnarray}
        where in principle, $k$ can be either real or imaginary and $u(n,k)$ is a four-component vector function of $n$ and $k$.
        Now, we impose the following geometry constraints: assume the system has a finite length $(0,N)$ and four-component eigenstates of $h(-i\partial_x)$, $\varphi_i(x)$, $i=1\ldots 4$, satisfy the following boundary conditions 
        \begin{eqnarray}
        \label{boundcond}
        \varphi_1(0)-\varphi_2(0)&=&\varphi_3(0)-\varphi_4(0)=0
        \label{boundary_condtion_0} \nonumber\\
        \varphi_1(N)+\varphi_2(N)&=&\varphi_3(N)+\varphi_4(N)=0 \label{boundary_condtion_1}.
        \end{eqnarray}
        These boundary conditions are justified in Appendix \ref{low-energy sector}.  We find that the eigenstates of $h(-i\partial_x)$, which also satisfy the boundary conditions (\ref{boundcond}), are characterized by
        quantized $k$ and $n$. The quantized values of $k$ are given by the following equation
        \begin{eqnarray}
        \tan{kN}=k/m, \label{quantized}
        \end{eqnarray}
        
        For future consideration, we introduce a notation $Q(m)$ representing the set of quantized values of $k$, ranging from $0$ to $\pi$, which satisfies Eq.~(\ref{quantized}). $\pi$, as the bound of quantized $k$, comes from that we consider unit lattice spacing and discrete space.  We will also use the notation $|\psi_{n,k}\rangle$ to represent a single MFT state, which is characterized by quantized $k$ and $n$. Its analytical expression in the coordinate space $x$, $\langle x |\psi_{n,k}\rangle$, is given in Appendix \ref{low-energy sector}.
        
        The other set of Majorana zero-energy boundary modes, which are protected by the topological number of the higher band, cannot be accurately obtained from the first-order expansion of the Hamiltonian (\ref{low_energy_ladder}). 
        Instead, one needs to use the exact form of the Hamiltonian, which can give a precise description of the higher-band. Solving $h(-i\partial_x)\psi(x)=0$, we find two zero-energy wave functions: $\psi_{L}(x)$ and $\psi_{R}(x)$, which are protected by the higher-band. The wave functions $\psi_{L,R}(x)$ are localized at left or right boundaries of the system correspondingly. Their analytical expressions read as
     \begin{eqnarray}
     \psi_L(x)&=&\sqrt{1-e^{-2/\xi}}e^{-x/\xi} \varphi_L\nonumber\\
     \psi_R(x)&=&\sqrt{1-e^{-2/\xi}}e^{-(N-x)/\xi}\varphi_R \label{Majorana_modes}
     \end{eqnarray}
     with $\xi^{-1}=\log \frac{2t}{\mu-t_v}$, $\varphi_L=\frac{1}{2}(1,1,-1,-1) ^T$ and  $\varphi_R=\frac{1}{2i}(1,-1,-1,1) ^T$. We remind the reader that $\xi^{-1}$ obtains the complex phase $i\pi$ if $2t/(\mu-t_v)<0$. We will use the  notation $| \psi_{L,R} \rangle$ to represent the corresponding ket (bra) states of these boundary modes.
     One may see that the expression for $\xi$ here is consistent with $\xi_-$ in Eq.(\ref{correlation_length}), which is the localization length derived from the lattice model. 
        
    \subsection{Low-energy effective matrix in the presence of the TRS breaking}
        
        Here we introduce the symmetry breaking field $B$ and study the low-energy properties of the TRS broken Hamiltonian. For example, we consider the Model I with the complex vertical pairing potential. We remind that the symmetry breaking term in this Hamiltonian is written in the momentum space as 
        \begin{eqnarray}
        h_{B}&=&-B \sigma_x \otimes \tau_y
        \end{eqnarray}
        with $B=\Delta^I_v$ in this case. 
        We confine our focus on the low-energy sector at the tricriticality. We choose the eigenstates of the parent Hamiltonian as basis states for the low-energy sector: $\{ |\psi_{L,R}\rangle,|\psi_{n,k}\rangle |n=\pm, k\in Q(m) \}$, where $Q(m)$ is set of quantized $k$-values defined above.
        In order to represent the full Hamiltonian, $h(-i\partial_x)+h_B$ in this basis, we need to calculate the corresponding matrix elements. Here $h(-i\partial_x)$ yields a diagonal matrix, whose diagonal values are the energies of corresponding modes. Thus, we will focus on the representation of $h_{B}$, which yields off-diagonal elements. 
        
        Upon calculating all the matrix elements of $h_B$, we find that only a limited number of off-diagonal matrix elements are non-zero. All possible finite off-diagonal elements are listed below. For the case when MFT-states $|\psi_{n,k} \rangle$ with real quantized $k$  couple to the zero-energy modes $\psi_{L,R}$ via $h_B$, we have         
        \begin{eqnarray}
        \langle \psi_{n,k} |h_{B}|\psi_{L}\rangle&=&-n \frac{B}{\sqrt{N}} \sqrt{\coth (1/2\xi)} \left|\sin kN \right|,   \label{off_1}\nonumber\\ 
        \langle \psi_{n,k} |h_{B}|\psi_{R}\rangle&=& \frac{B}{i\sqrt{N}} \sqrt{\coth (1/2\xi)}  \sin kN,
        \end{eqnarray}
        where $\left |\sin kN \right|$ means the absolute value of $\sin kN $. In the case with imaginary quantized $k=i\kappa$, one has
        \begin{eqnarray}
        \langle \psi_{n,i\kappa} |h_{B}|\psi_{L}\rangle&=&-i \sqrt{w} \frac{B}{\sqrt{N}} \sqrt{\coth (1/2\xi)},\nonumber\\
        \langle \psi^n_{n,i\kappa} |h_{B}|\psi_{R}\rangle&=&-n\sqrt{w} \frac{B}{\sqrt{N}} \sqrt{\coth (1/2\xi)} \label{off_2}.
        \end{eqnarray}
        Using the four matrix-elements above, we can calculate the corrections to all single particle energies due to the TRS breaking perturbation. From here one clearly observes the emergence of the  scale  
$g=\alpha(\xi)\sqrt{N} B$,
    with $\alpha(\xi)=\coth (1/2\xi)$ for Model I. This consideration analytically shows the nature of the scale $g$, which was extracted numerically from the universal finite-size corrections to the energy.  More details of determining $\alpha(\xi)$ for different models are contained in Appendix (\ref{alpha}).
        
        \subsection{Evolution of the low-energy spectrum with the TRS breaking field} \label{notation}
        \begin{figure*}
            \centering
            \subfloat[][$w=2$ ($4$-MF and $0$-MF)]{\includegraphics[scale=0.3]{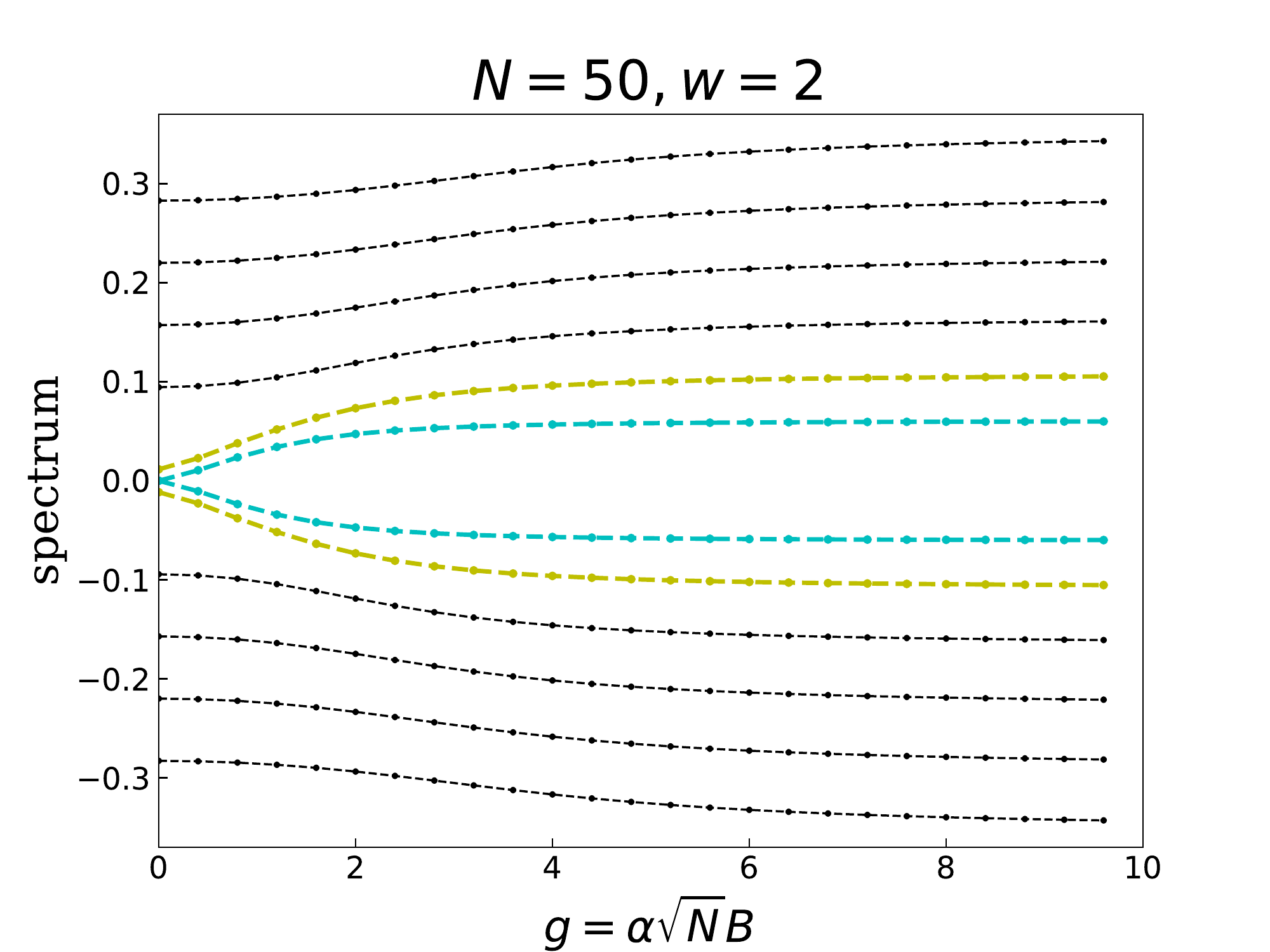}\label{spectrum_4M}}
            \subfloat[][$w=0$ (criticality)]{\includegraphics[scale=0.3]{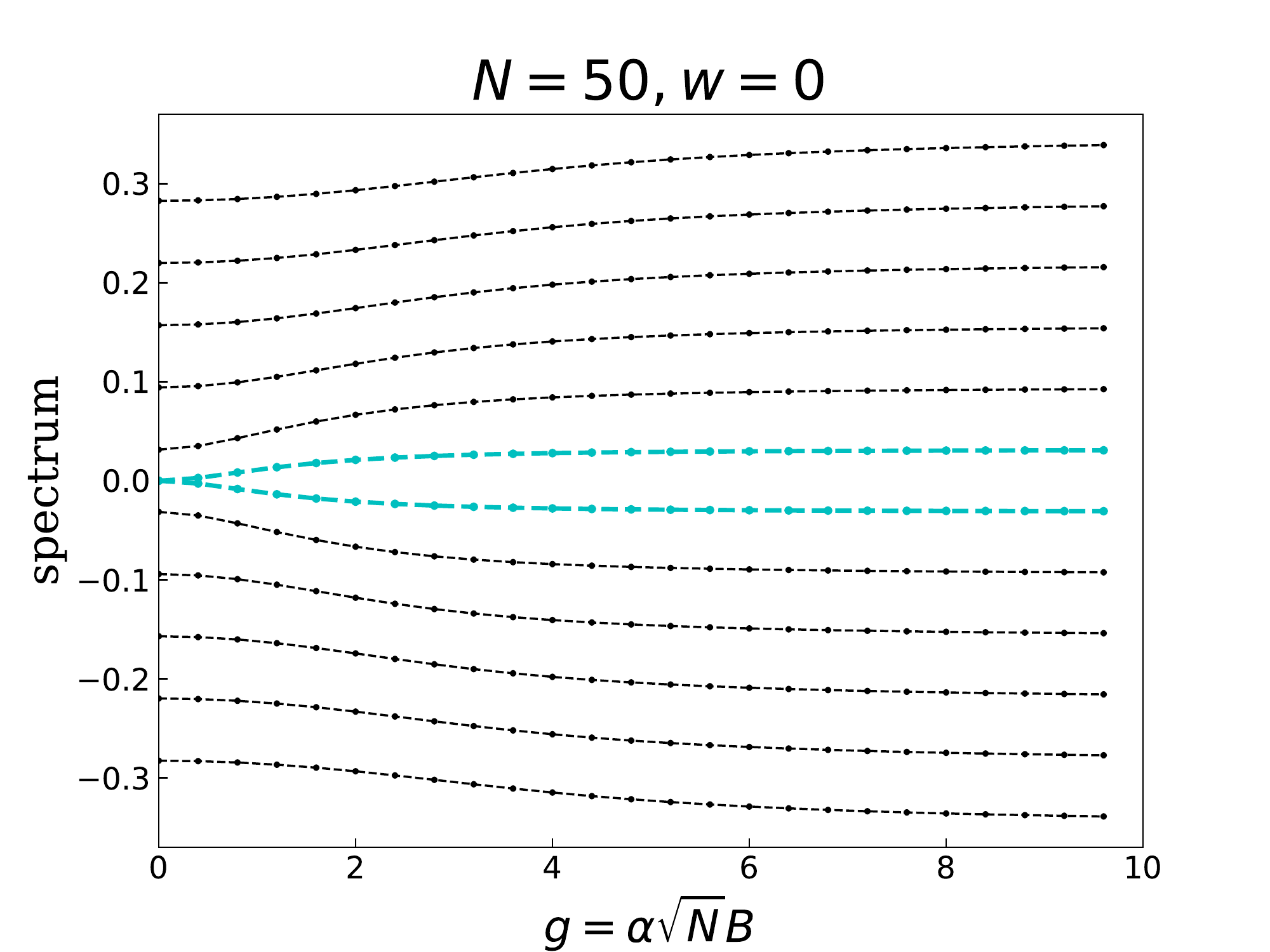}\label{spectrum_g}}
            \subfloat[][$w=-2$ ($2$-MF)]{\includegraphics[scale=0.3]{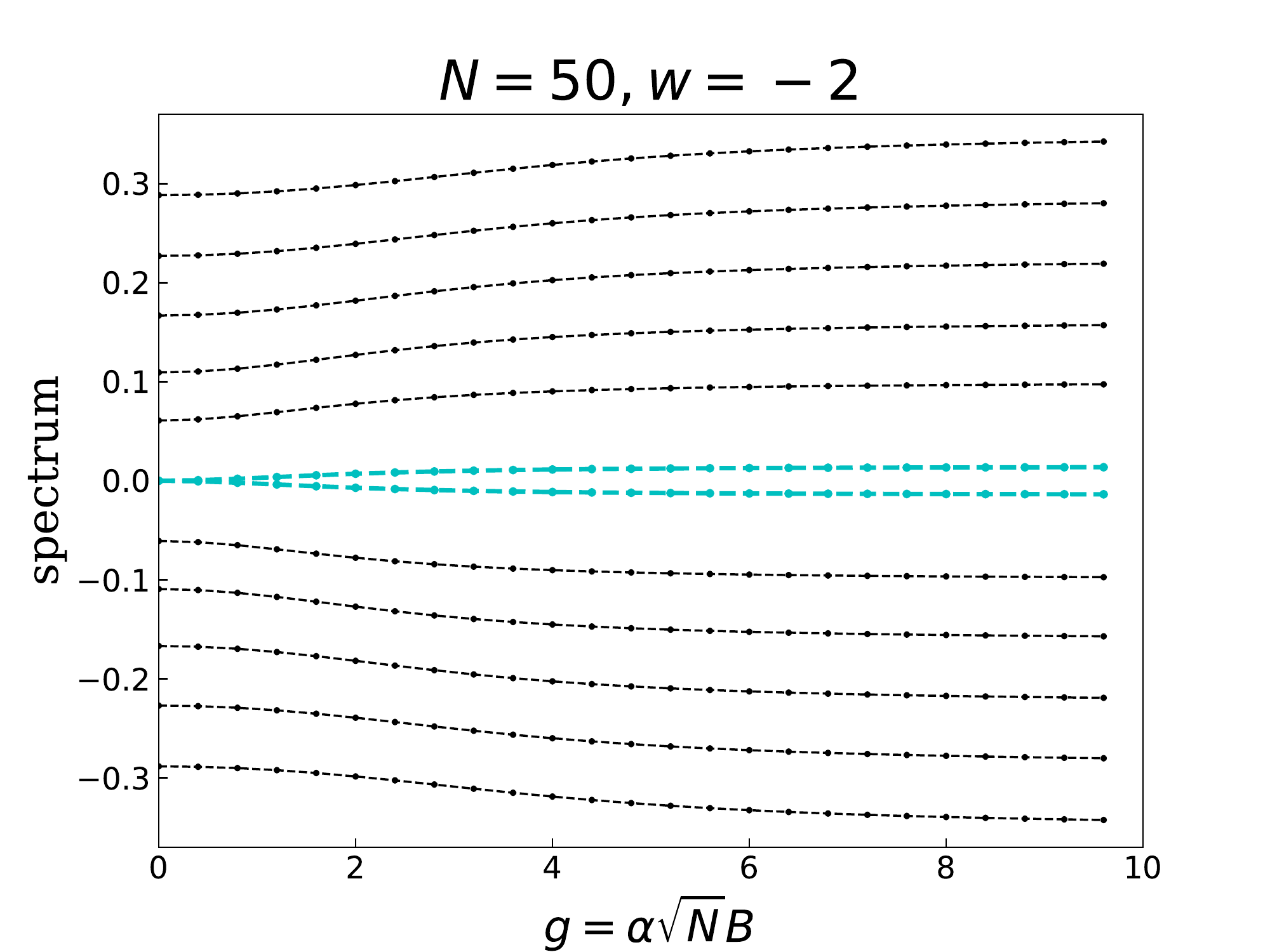}\label{spectrum_2M}}
            \caption{(Color online) Several low-lying energy levels of the effective matrix around $E=0$ plotted as a function of $g$ varying from $0$ to $10$. 
            Fig.~\ref{spectrum_4M} depicts the spectrum for $w=2$. It describes how $4$-MF state is smoothly connected to $0$-MF state. The evolution of the lowest energy levels, depicted as light gray (blue and yellow) lines,  shows how the zero-energy modes merge into the bulk. The bulk energy levels (black lines) also experience change. 
            Fig.~\ref{spectrum_g} depicts the spectrum for $w=0$ (criticality). It shows non-trivial behavior: the edge-modes, shown in light grey (blue), obtain finite energy and finally merge into the bulk. The energy levels of bulk modes deviate from Majorana CFT values when $g$ increases from $0$, and then converge to Majorana CFT values when $g\rightarrow +\infty$. Fig.~\ref{spectrum_2M} depicts the spectrum for $w=-2$. The zero-energy modes in $2$-MF obtain finite energy, which is proportional to $e^{-\lambda |w|}$, indicating that the localization length of edge modes changes.  }
            \label{xxxx}
        \end{figure*}
        One can calculate the eigenvalues of the effective low-energy matrix to obtain the low-energy spectrum of the system. We do this for three different values of $w=Nm$: (i)
        At $w=2$ we trace the topological transition from $4$-MF phase (at $g=0$) to $0$-MF phase (at finite $g$); (ii)
        At $w=0$ we trace the topological transition from tricriticality, (at $g=0$), which represents an example of gapless SPT phase with two boundary Majorana modes, to the trivial critical phase; (iii) At $w=-2$ we trace evolution of the spectrum of the topologically ordered $2$-MF phase with the TRS breaking field and show that the localized Majorana modes remain localized even in the presence of finite $g$. The low-energy spectra of the model in these three situations are shown in Fig.~(\ref{xxxx}):
        \begin{itemize}
            \item Fig. (\ref{spectrum_4M}) describes the evolution of the low-energy spectrum with scale $g$, when the $4$-MF phase smoothly transitions to the $0$-MF phase without closing the bulk gap. We clearly observe that four zero-energy modes get gapped and merge into the bulk band. At the same time, the energy levels of bulk modes also acquire a sensitivity to $g$. We will analyze their $g$-dependence afterward. 
            \item Fig.(\ref{spectrum_g}) describes the evolution of the low-energy spectrum starting from tricriticality at $g=0$, and upon increasing of $g$. We see that the zero-energy modes merge into the bulk band, and the energy level of each of the bulk mode is lifted by one "step" up, occupying the spot of the next higher band at $g=0$. This happens when $g$ increases from $0$ to $10$.
            \item Fig.~\ref{spectrum_2M} describes the $g$-dependence of the low-energy spectrum of the topologically ordered $2$-MF state.  Although the topology of $2$-MF phase is robust with respect to the  TRS breaking perturbation, the spectrum still exhibits some non-trivial behavior. The degeneracy of the zero-energy modes in $2$-MF phase is being slightly lifted, but the discrepancy between the states is exponentially small and is proportional to $e^{-|w|}$. In fact, the localization length $\xi$ of the edge modes is changed: if for example we consider the tricriticality corresponding to the upper tricritical point if the phase diagram \ref{pairingphase} (which has the coordinates $(0,1)$ in that diagram), then $\xi$ changes from $\xi_- (g=0)$ to $\xi_+ (g=\infty)$. We note that $g=0$ when the TRS is preserved ($B=0$), while $g\rightarrow \infty$ is achieved at the thermodynamic limit when the system size is $N\gg 1/(\alpha B)^2$). 
        \end{itemize}
        
        In Sect.(\ref{sector}) we considered the low-energy sector at the tricriticality ($g=0$) and showed it is composed of MFT states, and the zero-energy modes at the tricriticality are protected by the higher-energy band. Thus the low-energy spectrum at the tricriticality, which is labeled by $E(m,g)$, is given by
        \begin{eqnarray}
        E(m,0)=\pm\left\{0, \sqrt{m^2+k^2}| k\in \{Q_1(m)\ldots Q_N(m)\}\right\}. \nonumber\\ \label{g_0}
        \end{eqnarray}
        Here we observe that the low-energy spectrum away from tricriticality ($g\rightarrow \infty$, or, in practice, $g\gtrsim 10$) is given by
        \begin{eqnarray}
        E(m,g)=\left\{\pm \sqrt{m^2+k^2}| k\in \{Q_1(-m)\ldots Q_N(-m)\} \right\}. \nonumber\\ \label{g_1}
        \end{eqnarray}
        Here the spectral gap, $m$, around the tricriticality is positive in the $4$-MF and $0$-MF phase, while $m$ is negative in the $2$-MF phase.
        
        From Eq.(\ref{g_0}) and Eq.(\ref{g_1}), one can observe that for the same value of $m$, the set of quantized momenta, $k$, is reversed ($\{Q_1(m)\ldots Q_N(m)\}$ becomes $\{Q_1(-m)\ldots Q_N(-m)\}$) when g evolves from $0$ to $\infty$. The fact is supported by the numerically calculated energy spectrum shown in Figs.~\ref{spectrum_4M} and \ref{spectrum_2M}. Although at the criticality $Q(-0)=Q(0)$, Fig.~\ref{spectrum_g} still yields a non-trivial interpolation between $E(m=0,g=0)$ and $E(m=0,g=\infty)$, indicating the discussed in Sect.~(\ref{finite}) non-trivial finite-size scaling effect.
        
        The derivation of the closed analytical expression for the exact spectrum of the effective matrix is tedious since the edge modes $\psi_{L/R}$ couple to all other states. For small $g\ll 1$, one may use the quasiparticle picture at $g=0$ and apply perturbation theory. We find that the energies of MFT-states corresponding to the bulk $|\psi_{n,k}\rangle$ states, and the edge-modes $|\psi_{L,R}\rangle$, acquire perturbative corrections proportional to powers of the scale $g$. These corrections are discussed in detail in Appendix.(\ref{analytical}). In the limit $g\ll 1$ and $w=0$, the energy,  $\epsilon_e$, of the edge modes $\psi_{L/R}$ becomes
        \begin{eqnarray}
        \epsilon_e &\approx& g^2/N, |w|= 0
        \end{eqnarray}
        
        For the aymptotic region where $g$ and $w$ satisfy $\sqrt{|w|}e^{-|w|} \ll g \ll 1  $, $\epsilon_e$ is given by
        \begin{eqnarray}
        \epsilon_e &\approx&
        \begin{cases}
        \frac{\sqrt{2}}{|w|^3}g^2/N &\text{for $w\ll -1$}\\
        \sqrt{2w} g/N&\text{for $w\gg $1}
        \end{cases}
        \end{eqnarray}
        At $w\gg1$, the linearity coefficient $\sqrt{2w}\gg 1$. This implies 
        that the edge modes in $4$-MF phase are unstable with respect to the  TRS breaking perturbation. On contrary, at $w \ll -1$, the coefficient $\sqrt{2}/|w|^3\rightarrow 0$, indicating stability of edge modes in the $2$-MF phase.

        \subsection{The finite-size scaling function at $|w|, g\ll 1$}
        The ground state energy at $|w|\ll 1$ and $g\ll 1$ is obtained upon performing the summation of energy levels in the occupied band. At $|w|\ll 1$, the quantization of momenta in the MFT yields only bulk modes (and no boundary modes associated with MFT), and the energies of bulk MFT states are 
        $\{\epsilon_l|0\leq l\leq N-1\}$. There are, however, two Majorana edge modes protected by the topological index of the higher band.
       The energies of these edge modes (defined as $\psi_{L/R}$ in the previous section) are defined as $\pm\epsilon_e$.  The analytical expressions of $\epsilon_e$ and $\epsilon_l$ are given in Appendix.(\ref{analytical}). The ground state energy of the system will thus read 
        \begin{eqnarray}
        E_G&=&-\frac{1}{2} \epsilon_e-\frac{1}{2}\sum_{l\geq 0} \epsilon_l. \label{gs}
        \end{eqnarray}
This expression can further be simplified by rewriting the summation in the ground state energy (both, the summation in $\epsilon_e$ and the summation over $l$ in Eq.~(\ref{gs})) via a Cauchy contour integral. This is done in Appendix.~(\ref{analytical_fini_app}), leading to
\begin{eqnarray}
E_G&=&-\frac{1}{4}\oint_C \frac{dk}{2\pi i}[E_{m,g}(k)q_+(k)+E_{m,0}(k)q_-(k)]. \label{ge}
\end{eqnarray}
Here $q_\pm(k)=~\partial_k\ln(\cos\frac{1}{2}(Nk+\delta_m)\pm\frac{\pi}{4})$, $\tan \delta_m=k/m$, $E_{m,g}(k)~\approx\sqrt{m^2+k^2+8g^2/N^2}$ (since here $|w|\ll1$). The contour $C$ encircles all poles of $q_\pm(k)$ in the complex plane, but it avoids the branch cut line of $E(m,g)$ along the complex line $z=ix$ with $x\geq \sqrt{m^2+8g^2/N^2}$.

Upon taking the derivative of $q_\pm(k)$, one can obtain the bulk energy, $N\epsilon$, and the boundary energy, $b$, as follows:
\begin{eqnarray}\nonumber
N\epsilon&=&\int_{-\pi}^{\pi}\frac{dk}{2\pi}[E_{m,g}(k)+E_{m,0}(k)]N/4\\
b&=&\int_{-\pi}^{\pi}\frac{dk}{2\pi}[E_{m,g}(k)+E_{m,0}(k)]\partial_k\delta_m/4.
\end{eqnarray}
Then the combination $E_G-N\epsilon-b$ is the energy responsible for  finite size corrections to the ground state energy, $-f/N$. After performing a contour integration, we find the finite size scaling function, $f(w,g)$, given by the following integral: 
\begin{eqnarray}\nonumber
f(w,g)&=&-\int_{\sqrt{8g^2+w^2}}^{\infty}\frac{dz}{2\pi}E_{w,g}(z)\partial_z\ln (1+e^{-2z-2\delta_w(z)})\\
&-&\int_{w}^{\infty}\frac{dz}{2\pi}E_{w,0}(z)\partial_z\ln (1+e^{-2z-2\delta_w(z)}). \label{finite_analytical}
 \end{eqnarray}
        So, we see that $f(w,g)$ is a function of $w$ and $g$. 
        One can evaluate this integral at $w=0$ and $g\ll1$, yielding the following asymptote 
        \begin{eqnarray}
        f(0,g)-f(0,0)\simeq\frac{1}{\sqrt{\pi}}g^2 \log g,
        \end{eqnarray}
        where $f(0,0)=\pi/24$. 
        
        \section{The low-energy Hamiltonian at criticality and the boundary entropy}\label{boundary_Ising}
        In this section, we work out the second quantized formalism for the low energy Hamiltonian at criticality. We show that the boundary entropy is a universal function of the scale $g$ for Models I, II, III.
        \subsection{The low-energy Hamiltonian at criticality}
        In second quantization, for the MFT modes, $\psi_{n,k}$ and $\psi_{L/R}$ given by Eq.(\ref{MFT_modes}) and Eq.(\ref{Majorana_modes}), one defines the corresponding quasiparticle operators, $\hat{\psi}_{n,k}$ and  $\hat{\psi}_{\text{L/R}}$. The analytical expressions of these operators, and their properties are discussed in Appendix~(\ref{quasi}). The particle-hole symmetry of the theory is reflected in the following property of $\hat{\psi}_{n,k}$: $\hat{\psi}^\dagger_{+,k}=\hat{\psi}_{-,k}$. Thus, we only keep one operator with positive energy,  $\hat{\psi}_{k}\equiv \hat{\psi}_{n,k}$. We remind that $\hat{\psi}_{\text{L}}$ and $\hat{\psi}_{\text{L}}$ are two localized Majorana fermion operators:  $\hat{\psi}^\dagger_{\text{L}}=\hat{\psi}_{\text{L}}$ and $\hat{\psi}^\dagger_{\text{R}}=\hat{\psi}_{\text{R}}$.
        
      Consider the effective matrix corresponding to the Model I at the criticality ($m=0$), see Sec.(\ref{lowformt}). The corresponding low-energy Hamiltonian is given by
        \begin{eqnarray}
        H&=&\sum_{k\in Q(0)} k \hat{\psi}^\dagger_{k}\hat{\psi}_{k} +\frac{\alpha(\xi)B}{\sqrt{N}}\left( \hat{\psi}_{k}-\hat{\psi}^\dagger_{k} \right) \hat{\psi}_{\text{L}} \nonumber \\
        &+&\sin(kN)\cdot\frac{\alpha(\xi)B}{\sqrt{N}} \frac{\hat{\psi}_{k}+\hat{\psi}^\dagger_{k} }{i} \hat{\psi}_{\text{R}}  \label{B_low_Hamiltonian}
        \end{eqnarray}
        where $Q(0)\equiv\{ \pi/2N+m\pi/N |m=0,1,\dots, N-1\}$. Here the upper bound for $m$ is $N-1$, since the lattice spacing is chosen to be unity.
        
        In the continuum limit, when the lattice spacing $a\rightarrow0$, the Hamiltonian above becomes equivalent to (see Appendix (\ref{low_energy_critical}))
        \begin{eqnarray}
      H&=&\frac{1}{2}\int_{-\infty}^{+\infty} dx \begin{pmatrix}
      \hat{\psi}_c^\dagger(x) & \hat{\psi}_c(x)
      \end{pmatrix}(m(x)\sigma_z-i\sigma_y \partial_x)
      \begin{pmatrix}
      \hat{\psi}_c(x)\\
      \hat{\psi}_c^\dagger(x)
      \end{pmatrix}\nonumber\\
      &+& \frac{\alpha(\xi) B}{\sqrt{2}} \left (i\left[ \hat{\psi}_c^\dagger(0)+\hat{\psi}_c(0)\right] \hat{\psi}_{\text{L}} + \left[ \hat{\psi}_c^\dagger(N)- \hat{\psi}_c(N)\right] \hat{\psi}_{\text{R}} \right). \nonumber\\ \label{boundary_H}
      \end{eqnarray}
      Here $\hat{\psi}_c(x)$ is the generic spinless fermion annihilation operator at continuous space coordinate $x$, satisfying $\{\hat{\psi}_c(x),\hat{\psi}_c^\dagger(x') \}=\delta(x-x')$. Operators $\hat{\psi}_{\text{L/R}}$ correspond to two localized Majorana fermion operators, satisfying $2\hat{\psi}^2_{\text{L/R}}=1$, and the mass term, $m(x)$, is given by
     $$ m(x)=\left\{
  \begin{array}{rcl}
  0       &      & 0<x<N\\
  +\infty     &      & x<0, x>N
  \end{array} \right.  \label{mass_file}
   $$
  From Eq.(\ref{boundary_H}), one may formulate the corresponding action and conclude that $\alpha(\xi)B$ drives the boundary RG flow from the bulk Ising criticality with two localized Majorana boundary modes (gapless SPT phase) to the bulk Ising criticality without localized Majorana boundary modes (gapless trivial phase). Scaling dimension of $\alpha(\xi)B$ is $1/2$ so the dimensionless $g=\alpha(\xi)\sqrt{N}B$ emerges as invariant scale under RG flow. The Hamiltonian in Eq.(\ref{boundary_H}) is consistent with the action proposed in the literature\cite{prb17Cornfeld,prb92Emery}, which is used to describe the boundary Ising chain. In the model of boundary Ising chain , boundary magnetic field drives the RG flow from free boundary condition\cite{npb89Cardy,npb91Cardy} to fixed boundary condition.
  
  \subsection{The boundary entropy}
The boundary entropy is defined by the logarithm of universal non-integer degeneracy\cite{prl91Affleck}, which only depends on the universality class of the conformally invariant boundary condition\cite{npb89Cardy,npb91Cardy}. In practice, one can obtain the boundary entropy via calculating the Von-Neumann entropy\cite{jop09Cardy}. Given the system defined on an interval $M$, one can partition it into two subsystems consisting of two intervals, $A$ and $B$. Von-Neumann entropy measures the entanglement between two subsystems. Let $\rho$ be the density matrix of system $M$ and $\rho_A :=\text{tr}_{B} \rho$ is the reduced density matrix of $A$. Von-Neumann entropy is then defined as $S=-\text{tr}_{A} \rho_A \log \rho_A $. In the following, we will focus on the system with boundaries: $M=(0,N)$, $A=(0,l)$ and $B=(l,N)$.

\begin{figure}
        \includegraphics[scale=0.5]{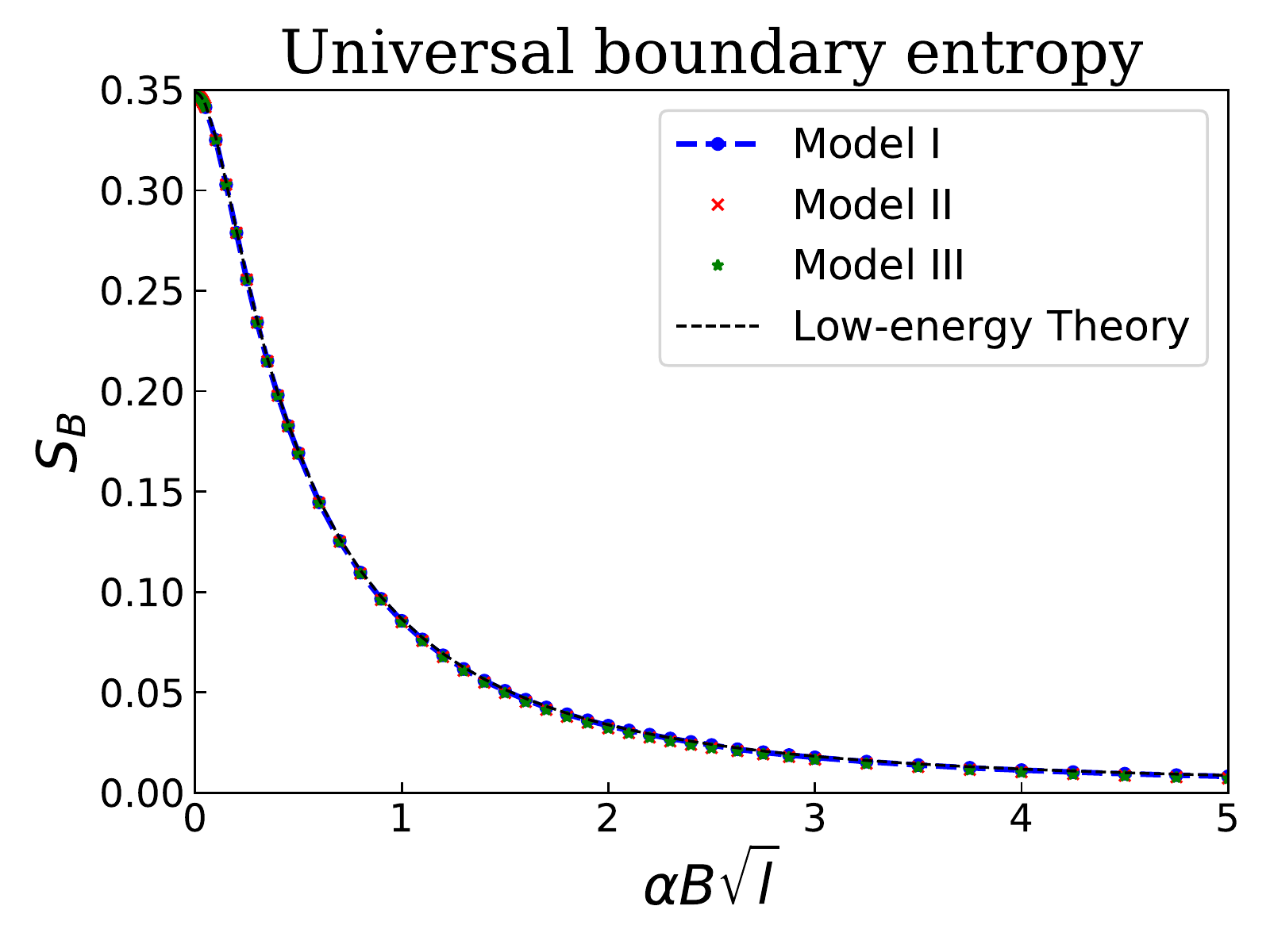}
        \centering
        \caption{(Color online) Boundary entropies of Models I, II, III and the low-energy effective Hamiltonian  Eqs.(\ref{B_low_Hamiltonian}) and Eq.(\ref{boundary_H}). All curves fall into same universal curve representing the universal boundary entropy. 
        } 
        \label{boundary_entropy}
    \end{figure}

For critical lines around the tricriticality, we find that the von-Neumann entropy is given by 
     \begin{eqnarray}
     S=\frac{c}{6} \log(2l)+c_1/2+S_B(\alpha B \sqrt{l})
     \end{eqnarray}
 where $c_1$ is non-universal constant and $S_B(\alpha B \sqrt{l})$ is the boundary entropy. We show that $S_B(\alpha B \sqrt{l})$ is universal for Models I, II, III and the low-energy Hamiltonian Eq.(\ref{boundary_H}). It is depicted in Fig.(\ref{boundary_entropy}). This result is consistent with that reported in the literature on the universal flow of boundary entropy\cite{iop09Affleck,prb06Zhou,prb17Cornfeld}.
  
    

        \section{conclusions} \label{conclusion}
        In this paper, we report the existence of quantum tricriticality in Models I, II, and III, which separates topologically ordered, SPT, and trivial phases of the system. We study the finite size corrections to the ground state energy and find that it is a universal function of a new dimensionless scale, $g=\alpha \sqrt{N}B$, where $B$ is the symmetry-breaking field. Thus, around the tricriticality, the finite size corrections are determined by two variables, $w=Nm$, and $g$. We derive the effective low-energy theory corresponding to these models and show the emergence of the scale $g$ that describes the evolution of the low-energy spectrum with $B$. We analytically calculate the asymptotes of the universal finite-size scaling function, $f(w,g)$. Finally, we show that the scale $g$ emerges  in the boundary entropy, which is shown to be universal for three models.
        
        We conjecture that the universal finite-size double-scaling function should emerge not only for coupled Majorana chains considered in this work, but also for other models which support SPT, topologically ordered, and trivial phases. In the free fermion classification\cite{pu01Kitaev,prb08Schnyder,rmp16Chiu}, only Majorana chains can support such phases. However, one may find such tricriticalities in interacting models, including the spin chains, which are beyond the free fermion classification. For example it is interesting to investigate the finite size effects in  
        critical spin chains perturbed by a TRS breaking scalar chirality operator\cite{npb01sedrkyan,ahp06Mkhitaryan}.  
        The tricriticality may also exist in interacting Majorana chains \cite{prb10Fidkowski,prb15Katsura,prb15Lang,prb16Herviou,prl15Iemini,prb11Fidkowski},since $4$-MF and $2$-MF are both stable in the presence of interactions, which preserve the time-reversal symmetry and fermion-parity symmetry. The behavior of finite-size scaling function in interacting systems, including the spin chains and higher dimensional systems with TRS breaking is beyond the scope of the paper, and is left for future studies.

        \section{Acknowledgements} 
        The research was supported by startup funds from UMass Amherst. 
    
        \appendix
        \section{Zero-energy modes of antisymmetric matrix $A$ } \label{Majorana_zero_modes}
        In the Majorana-fermion basis, one can formulate the Hamiltonian $H$ in Eq.(\ref{total_hamiltonian}) in terms of an anti-symmetric matrix, $A$,
        \begin{eqnarray}
        H&=&\frac{i}{4} \sum^2_{m,n}\sum^{2N}_{j,l}c_{j,m} A^{m,n}_{j,l} c_{l,n}
        \end{eqnarray}
        where the matrix elements are given by: $~A^{m,m}_{2j-1,2j}=\mu$, $~A^{m,m}_{2j,2j+1}=\Delta+t$, $~A^{m,m}_{2j-1,2j+2}=\Delta-t$ where $m=1,2$, $~A^{1,2}_{2j-1,2j}=\Delta_v-t_v$, $~A^{1,2}_{2j,2j-1}=\Delta_v+t_v$.
        Then the problem of solving for the edge modes with open boundary conditions is equivalent to find new Majonara modes\cite{pu01Kitaev} $b=\sum_{j,m} v_{j,m} a_{j,m}$ where $\vec{v}$ satisfies
        \begin{eqnarray}
        \sum_{j,m} v_{j,m} A^{m,n}_{jl}=v_{l,n} \cdot 0  \label{edgeequ}
        \end{eqnarray}
        In the case when $\Delta=t$, one can further simplify Eq.(\ref{edgeequ}), leading to series of equations, which in turn provide two types of solutions. { One solution for  $v_{j,m}$ is such that only $v_{j,m}$ with odd site indices $j$ are non-zero:
        \begin{eqnarray}
        v_{2i-1,m}=\alpha_{+,m} x^{-i}_++\alpha_{-,m} x^{-i}_-
        \end{eqnarray}
        The other solution is such that only $v_{j,m}$ with even site indices $j$ are non-zero:
        \begin{eqnarray}
        v_{2i',m}=\beta_{+,m} x^{-i'}_++\beta_{-,m} x^{-i'}_-
        \end{eqnarray}
        where  $i$ acquires values from $1,\ldots , N$, while  $i'=N-i$.} Here the coefficients $\alpha$ and $\beta$ are arbitrary. Moreover, $x_+$ and $x_-$ are given by
        \begin{eqnarray}
        x_{\pm}=\frac{2t}{\mu\pm \sqrt{t^2_v-\Delta^2_v}} \label{edge}
        \end{eqnarray}
        
        The boundary conditions are $v^m_{2i-1}=0$ for $i=L+1$ and
        $v^m_{2i'}=0$ for $i'=L+1$. Thus the existence of edge modes reduces to the estimation of $x_\pm$: 
        \begin{itemize}
            \item  $x_+>1$ and  $x_->1$ indicate 4 localized Majorana edge modes
            \item  $x_+>1$ and  $x_-<1$  indicate 2 localized Majorana edge modes
            \item  $x_+<1$ and  $x_->1$  indicate 2 localized Majorana edge modes
            \item  $x_+<1$ and  $x_-<1$  indicate 0 localized Majorana edge modes.
        \end{itemize}
        One may convert  $x_{\pm}$ into localization length of the edge modes with the help of  $e^{-1/\xi_{\pm}}=(x_\pm)^{-1}$, which can be further written as 
        \begin{eqnarray}
        \xi^{-1}_{\pm}=\ln \frac{2t}{\mu\pm \sqrt{t^2_v-\Delta^2_v}}. \label{correlationlength}
        \end{eqnarray}
        Here $\xi_{\pm}$ are the correlation lengths of edge modes characterized by $x_\pm$, respectively.
        
        \section{Uniform flux} \label{uniform_app}
         In the momentum-space representation, $t$-dependent terms  are obtained upon Fourier transforming $(H_1+H_2)$ evaluated at $\Delta=\mu=0$. This yields the expression
        \begin{eqnarray}
        \nonumber
        \label{zero}
        &-&\sum_{j}( 2t \cos(k+\theta/2) \hat{a}^{\dagger}_{k,1}\hat{a}_{k,1}+h.c.)\\
        &-&\sum_{j}( 2t \cos(k-\theta/2) \hat{a}^{\dagger}_{k,2}\hat{a}_{k,2}+h.c.)
        \end{eqnarray}
        Now, we switch to the Majorana basis, and express the complex hopping as $te^{i\theta/2}=t_R+it_I$. This will 
        provide the following form of Eq.~(\ref{zero}) for $(H_1+H_2)|_{\Delta=\mu=0}$:
        \begin{eqnarray}
        &&    -\frac{it_I}{2}\sum_{j=1}^{N-1}\sum_{m=1}^{2} e^{im \pi}\left(c_{2j-1,m}c_{2j+1,m}+ c_{2j,m}c_{2j+2,m}\right) \nonumber\\
        &&+\frac{it_R}{2}\sum_{j=1}^{N-1}\sum_{m=1}^{2} \left(c_{2j,m}c_{2j+1,m}- c_{2j-1,m}c_{2j+2,m}\right)
        \end{eqnarray}
        
        \section{Staggered flux} \label{stagger app}
                Since \text{staggered-flux} introduce \text{sub-lattice}, $t^I_v$ gives coupling between two valleys in momentum-space
     In the Majorana basis, we decompose $t_v e^{i\theta/2}=t^R_v+it^I_v$ and express interchain Hamiltonian as
        \begin{eqnarray}
        \nonumber
        H_\text{interchain}&=&\frac{i}{2}\sum_j(\Delta_v-t^R_v)c_{2j-1,1}c_{2j,2}\\
        &+&\frac{i}{2}\sum_j(\Delta_v+t^R_v)c_{2j,1}c_{2j-1,2} \nonumber\\
        &+&\frac{i}{2}\sum_{j=2n+1}t^I_v(c_{2j-1,1}c_{2j-1,2}+ c_{2j,1}c_{2j,2}) \nonumber\\
        &-&\frac{i}{2}\sum_{j=2n}t^I_v(c_{2j-1,1}c_{2j-1,2}+ c_{2j,1}c_{2j,2}) 
        \end{eqnarray}
        where terms propotional to $t^I_v$ break \text{TRS}-symmtery.
        
        Staggering of the flux doubles the unit cell and introduces two sub-lattices. Hopping $t^I_v$ couples two valleys in momentum-space
        \begin{eqnarray} 
H_\text{interchain}(t^I_v, t^R_v=0,\Delta_v=0)\nonumber\\
=\sum_{k}it^I_v \hat{a}^\dagger_{k,1}\hat{a}_{k+\pi,2}+h.c.
        \end{eqnarray}
            
        \section{MFT of the low-energy sector}
    \label{low-energy sector}
            The operator $h(-i\partial_x)$ has two types of eigenstates: one is $e^{ikx}\psi_k$ where $\psi_k$ satisfies that 
        $h(k)\psi_k=E \psi_k$, representing the bulk modes. The other one is $e^{-\kappa x}\psi_{i\kappa}$, which correspond to edge modes.  Here we are only interested in the low-energy sections in the eigenspace of $h(k)$, and may adopt $\Delta_v=0$ and $w_v>0$ to simplify the representation of the wavefunction. There are two low-energy branches, one is positive and the other is negative:
      \begin{eqnarray}
      &h(-i\partial_x)&e^{ikx}u_{\pm}(k)=\pm 2tE_ke^{ikx}u_{\pm}(k). 
        \end{eqnarray}
        Here $E_k=\sqrt{m^2+k^2}$ and $m=\frac{2t-\mu-\sqrt{w^2_v-\Delta^2_v}}{2t}$ is the gap of the system. The definition of $m$ indicates $m>0$ for $4$-MF and $m<0$ for $2$-MF. 
        
        The wavefunctions $u_{\pm}(k)$, with real $k$, represent the bulk modes. They are given by
        \begin{eqnarray}
        u_{\pm}(k)&=&\frac{\sqrt{2}}{2}\begin{pmatrix}
        -\cos \theta_\pm&i \sin \theta_\pm&-\cos \theta_\pm&i \sin \theta_\pm
        \end{pmatrix}^T. \nonumber
        \end{eqnarray}
        Here  $\theta_\pm$ is defined as 
        \begin{eqnarray}
        (\cos \theta_\pm, \sin \theta_\pm)&=&\frac{(m\pm E_k,k)}{\sqrt{2E_k(E_k\pm m)}} 
        \end{eqnarray}
        The wavefunction with imaginary $k=i\kappa$ corresponds to boundary excitations:
        \begin{eqnarray}
        u_\pm(i\kappa)&=&\frac{\sqrt{2}}{2}\begin{pmatrix}
        \cos \phi_\pm& \sin \phi_\pm&\cos \phi_\pm& \sin \phi_\pm
        \end{pmatrix}^T.
        \end{eqnarray}
        The variables $\phi_\pm$ here are defined as 
        \begin{eqnarray}
        \label{phi}
        ( \cos \phi_\pm, \sin\phi_\pm)&=&\frac{(m\pm E_{i\kappa},\kappa)}{\sqrt{2m(m\pm E_{i\kappa})}}.
        \end{eqnarray}
        The boundary conditions in Eq.(\ref{boundary_condtion_0}) are obtained by imposing a chemcial potential, $\mu(x)$, such that
        \begin{eqnarray}
           \mu(x)=\left\{
  \begin{array}{rcl}
  \mu       &      & 0<x<N\\
  -\infty     &      & x<0, x>N .
  \end{array} \right. 
                \end{eqnarray}
    The chemical potential $\mu(x)$ imposes the condition that the mass, $m(x)=\left(2w-\mu(x)-w_v\right)/(2t)$, is $+\infty$ outside of the segment $(0,N)$.
    
        For each energy level, there exist two linearly-independent eigenstates, with $k$ and $-k$. To satisfy the boundary conditions, one chooses a linear superposition of two eigenstates, which provides an eigenstate that is consistent with Eq.~(\ref{boundary_condtion_0}).
        This yields a set of two
       normalized wavefunctions of the form
        \begin{eqnarray}
        \psi_{\pm,k}(x)&=&\frac{1}{{\sqrt{2N}}}\left({e^{i(kx-\theta_\pm)}} u_\pm(k)-{e^{-i(kx-\theta_\pm)}} u_\pm(-k) \right),  \nonumber\\
        \label{MFT_modes}\\
        \psi_{\pm,i\kappa}(x)&=&\sqrt{\frac{w}{N}}\left({e^{-\kappa x}} u_\pm(i\kappa)\mp{e^{\kappa (x-N)}} u_\pm(-i\kappa)\right)
        \end{eqnarray}
        where the normalization factors ignore the overlapping terms between $k$ and $-k$ since overlap bewtween wavefunctions at $k$ and $-k$ is highly-oscillatory, while the overlapping bewteen wavefunctions with $k=i\kappa$ and $k=-i\kappa$ is exponentially small (as $e^{-N\kappa}$). 
                Finally, the boundary conditions at $x=N$, which are given by Eq.(\ref{boundary_condtion_1}), yield the quantization rule of $k$ as $\tan{kN}=k/m.$
        \section{Quasiparticle operators}\label{quasi}
        We define the fermion creation operators $\hat{\psi}^{\dagger}_{n,k}$ and $\hat{\psi}^{\dagger}_{\text{L/R}}$ as follows
     \begin{eqnarray}
 \hat{\psi}^{\dagger}_{n,k}&=&\sum_{i=1}^{N} \begin{pmatrix}
 \hat{a}^\dagger_{i,1}&\hat{a}_{i,1}&\hat{a}^\dagger_{i,2}&\hat{a}_{i,2}
 \end{pmatrix}\cdot \psi_{n,k}(i-\frac{1}{2})\nonumber\\
 \hat{\psi}^{\dagger}_{\text{L}}&=&\sum_{i=1}^{N} \begin{pmatrix}
 \hat{a}^\dagger_{i,1}&\hat{a}_{i,1}&\hat{a}^\dagger_{i,2}&\hat{a}_{i,2}
 \end{pmatrix}\cdot \psi_{\text{L}}(i-1)\nonumber\\
 \hat{\psi}^{\dagger}_{\text{R}}&=&\sum_{i=1}^{N} \begin{pmatrix}
 \hat{a}^\dagger_{i,1}&\hat{a}_{i,1}&\hat{a}^\dagger_{i,2}&\hat{a}_{i,2}
 \end{pmatrix}\cdot \psi_{\text{R}}(i)
 \end{eqnarray} 
 where $\hat{a}_{i,j}$ is the fermion annihilation operator at lattice site $i$ and chain $j$,  $\psi_{n,k}$ and $\psi_{\text{L/R}}$ are the wavefunctions given by  Eq.(\ref{MFT_modes}) and Eq.(\ref{Majorana_modes}). One can show that the operators defined above satisfy  $\hat{\psi}^{\dagger}_{n,k}=\hat{\psi}_{-n,k}$ and $\hat{\psi}^{\dagger}_{\text{L/R}}=\hat{\psi}_{\text{L/R}}$.
 
 \section{The low-energy Hamiltonian in the continuum limit} \label{low_energy_critical}

Given the lattice constant, $a$, the lattice sites can be chosen to be located at points  
$X=\left \{ ja- a/2  |j=1,2\dots N/a \right\}$. Then, we consider $H_0$ as 
 \begin{eqnarray}
H_0=\sum_{x} \begin{pmatrix}
      \hat{\psi}^\dagger(x) & \hat{\psi}(x)
      \end{pmatrix}(m(x)\sigma_z-i\sigma_y \partial_x)
      \begin{pmatrix}
      \hat{\psi}(x)\\
      \hat{\psi}^\dagger(x)
      \end{pmatrix}\nonumber.
 \end{eqnarray} 
Here $\hat{\psi}(x)$ is the fermion annihilation operator at the lattice site, $x$, 
  with $x=ja- a/2$, $j\in \mathbb{Z}$ in the summation. Here $m(x)=0$ if $x\in X$ and $m(x)=-\infty$ if $x\notin X$.
 One may define an operator, $\hat{\psi}_k$, in terms of which $H_0$ is diagonal,
\begin{eqnarray}
\hat{\psi}_{k}=\frac{i}{\sqrt{N/a}}\sum_{x\in X}  \cos(kx-\pi/4) \hat{\psi}^\dagger(x)- \sin(kx-\pi/4) \hat{\psi}(x).\nonumber \\ \label{k_operator}
\end{eqnarray}
 Here $k= \frac{\pi}{N} (j-1/2), j=1,2\dots N/a $. Then the Hamiltonian, $H_0$, is  $H_0=\sum_k k \hat{\psi}^\dagger_{k}\hat{\psi}_{k}$. In fact, Eq.(\ref{k_operator}) can be rewritten to express $\hat{\psi}(x)$ as
 \begin{eqnarray}
\hat{\psi}(x)=\frac{i}{\sqrt{N/a}}\sum_{k}  \cos(kx-\pi/4) \hat{\psi}^\dagger_k+ \sin(kx-\pi/4) \hat{\psi}_k.\nonumber  \label{x_operator}
\end{eqnarray}
Then one can use the equation above to show the following identities
 \begin{eqnarray}
\frac{i}{\sqrt{2a}} \left(  \hat{\psi}(\frac{a}{2}) +\hat{\psi}^\dagger(\frac{a}{2})\right )=\frac{1}{\sqrt{N}} \sum_k \cos(\frac{ka}{2}) (\hat{\psi}_k-\hat{\psi}^\dagger_k), \nonumber
\end{eqnarray}
and 
\begin{eqnarray}
&&\frac{i}{\sqrt{2a}} \left(  \hat{\psi}(N-\frac{a}{2})+\hat{\psi}^\dagger(N-\frac{a}{2})\right) \nonumber\\
&=&\frac{1}{\sqrt{N}} \sum_k \sin(kN)\cos(\frac{ka}{2})  \frac{\hat{\psi}_k+\hat{\psi}^\dagger_k}{i}.
\end{eqnarray}
In the continuum limit ($a\rightarrow0$) one can introduce the field operator $\psi_c(x)= \lim_{a\rightarrow0}\frac{1}{\sqrt{a}} \psi(x)$ so that $\{\psi_c(x),\psi^\dagger_c(x')\}=\delta(x-x')$.  Since we are interested in the low-energy theory (and small $k$), we  approximate $\cos({ka}/{2}) \simeq 1$. Thus, one will obtain
\begin{eqnarray}
\frac{i}{\sqrt{2}} \left(  \hat{\psi}_c(0) +\hat{\psi}_c^\dagger(0)\right )&=&\frac{1}{\sqrt{N}} \sum_k  (\hat{\psi}_k-\hat{\psi}^\dagger_k) \label{relations}\\ 
\frac{1}{\sqrt{2}} \left(  \hat{\psi}_c^\dagger(N)-\hat{\psi}_c(N)\right) &=&\frac{1}{\sqrt{N}} \sum_k \sin(kN)  \frac{\hat{\psi}_k+\hat{\psi}^\dagger_k}{i}. \nonumber
\end{eqnarray}
Now we consider the Hamiltonian with  symmetry-breaking field, $B$, from Eq.(\ref{B_low_Hamiltonian})
\begin{eqnarray}
        H&=&\sum_{k} k \hat{\psi}^\dagger_{k}\hat{\psi}_{k} +\frac{\alpha(\xi)B}{\sqrt{N}}\left( \hat{\psi}_{k}-\hat{\psi}^\dagger_{k} \right) \hat{\psi}_{\text{L}} \nonumber \\
        &+&\sin(kN)\cdot\frac{\alpha(\xi)B}{\sqrt{N}} \frac{\hat{\psi}_{k}+\hat{\psi}^\dagger_{k} }{i} \hat{\psi}_{\text{R}} . \label{appendix_B_Hamiltonian}
        \end{eqnarray}
Here $k= \frac{\pi}{N} (j-1/2), j=1,2\dots +\infty $ due    to $a\rightarrow 0$. One can use Eqs.(\ref{relations}), to show that the Hamiltonian in Eq.(\ref{appendix_B_Hamiltonian}) is equivalent to Eq.(\ref{boundary_H}).
\section{The determination of model-dependent function $\alpha(\xi)$ }\label{alpha}
        When the localization length $\xi$ is finite, especially comparable/larger to lattice spacing, the nature of short-ranged symmetry-breaking Hamiltonian reflects itself as model-dependent function $\alpha(\xi)$ in the scale $g$. This section talks about how to determine this model-dependent function. 
        
        In practice, $\alpha(\xi)$ can always be found numerically for arbitrary lattice models. For the system known with the finite localization length $\xi$ (note $\xi$ is found at $B=0$, i.e., $\xi$ is independent of $B$ ), one can obtain the finite-size scaling function by tuning symmetry breaking field $B$ and plot the finite size scaling function $f$ versus $\alpha B \sqrt{N}$, where $\alpha$ is the parameter to be tuned. One can tune $\alpha$ until the plot is fitting well with the curves, for example, shown in Fig. (\ref{central}) and Fig. (\ref{uw2}). Then one can extract the value of $\alpha$ at a fixed value of the localization length, $\xi$.  
        
        Also, $\alpha(\xi)$ can be obtained analytically. In the Sect.(\ref{lowformt}), we show that $\alpha(\xi)=\sqrt{\coth(1/2\xi)}$ for Model I. It is calculated from the effective matrix elements of symmetry-breaking Hamiltonian, $h_B$. The method has also been applied to Model III. Symmetry-breaking Hamiltonian $h_B(x)$ in Model III is given by 
            \begin{eqnarray}
            h_B(x)=-B (-1)^x \mathbf{1}_2 \otimes \tau_y
            \end{eqnarray}
        One can show that the effective matrix elements $\langle \psi_{n,k} |h_{B}|\psi_{L} \rangle=-n \frac{B}{\sqrt{N}} \sqrt{\tanh (1/2\xi)} \left|\sin kN \right|$. Then one finds that $\alpha(\xi)=\sqrt{\tanh(1/2\xi)}$ for Model III. 
        However, finding $\alpha(\xi)$ for Model II analytically still remains an open problem.
        \section{Effective matrix} \label{Appendix}
        
        Around the tricriticality, we find matrix representation of $h=h(-i\partial_x)+h_B$ using the basis of $N+2$ normalized wavefunctions, $V=~\{\psi_{L},\psi_{R},\psi_{+,k_1},\psi_{-,k_1},..., \psi_{+,k_N},\psi_{-,k_N}\}\equiv\{v_i|i=1,2,\cdots,N+2\}$, where $k_2<...<k_N$ and $k_n\in Q_R$ for $n \geq 2$. Momentum $k_1$ could be either real or imaginary. Then the matrix representation of $h=h(-i\partial_x)+h_B$ gives raise to matrix elements $h_{ij}=\langle v_i |h|v_j\rangle$, which will be presented below.     However, before going through the details, we introduce a new variable, $x=kN$, $k\in Q(m)$ (we remind that $N$ is the size of the system). This definition implies that $x$ satisfies the equation     
        \begin{eqnarray}
        \tan x=x/w \label{x_quantize}
        \end{eqnarray}
        where $w=Nm$. Solutions to Eq.~(\ref{x_quantize}) are labeled by $x_l$, $0\leq l<N$.  The meaning of $x_l$ is clarified in the following: 
        i) At the crticiality and 2-MF phase ($w\leq 1$), all the solutions to the Eq.(\ref{x_quantize}) are real, then $x_l$ is given by $x_l=l\pi+\phi_l, 0\leq \phi_l<\pi$, $0 \leq l$. ii) At the 4-MF phase (w> 1), then there exists imaginary solution. So $x_l$ is given by $x_0=iq$, $0<q\leq w$ and $x_l=l\pi+\phi_l, 0\leq \phi_l<\pi$, $1 \leq l <N$.
        
        The diagonal part of $h$ is the same for all 4-MF, 2-MF and 0-MF phases. The diagonal matrix elements are
        \begin{eqnarray}
        h_{11}=h_{22}&=&0\\ 
        h_{2l+3,2l+3}&=&-h_{2l+4,2l+4}=\frac{\sqrt{w^2+x_l^2}}{N} 
        \end{eqnarray}
        where $l \geq 0$. 
                However, off-diagonal matrix elements are somewhat more complicated, due to different topological natures of phases. For 2-MF phase and the corrsponding critical points, the non-zero off-diagonal matrix elements are given by 
        \begin{eqnarray}
        h_{1,2l+3}=h_{2l+3,1}&=&-\frac{g}{N} \frac{x_l}{ \sqrt{w^2+x^2_l}}\label{m1} \nonumber\\ 
        h_{1,2l+4}=h_{2l+4,1}&=&\frac{g}{N} \frac{x_l}{ \sqrt{w^2+x^2_l}} \nonumber\\
        h_{2,2l+3}=h_{2l+3,2}&=&e^{i\pi l}\frac{g}{iN} \frac{x_l}{ \sqrt{w^2+x^2_l}}\nonumber\\ 
        h_{2,2l+4}=h_{2l+4,2}&=&e^{i\pi l}\frac{g}{iN} \frac{x_l}{ \sqrt{w^2+x^2_l}} \label{m4}
        \end{eqnarray} 
        For 4-MF phase, the matrix elements coincide with Eq.(\ref{m4}) except the following ones:
        \begin{eqnarray}
        h_{1,3}=-h_{3,1}&=&i \sqrt{w} \frac{g}{N}\nonumber\\ 
        h_{1,4}=-h_{4,1}&=&i \sqrt{w} \frac{g}{N} \nonumber\\
        h_{2,3}=-h_{3,2}&=& \sqrt{w} \frac{g}{N} \nonumber\\
        h_{2,4}=-h_{4,2}&=&- \sqrt{w} \frac{g}{N}. 
        \end{eqnarray}

    \section{Analytical spectrum at small $g$} \label{analytical}
            \begin{figure}
            \includegraphics[scale=0.5]{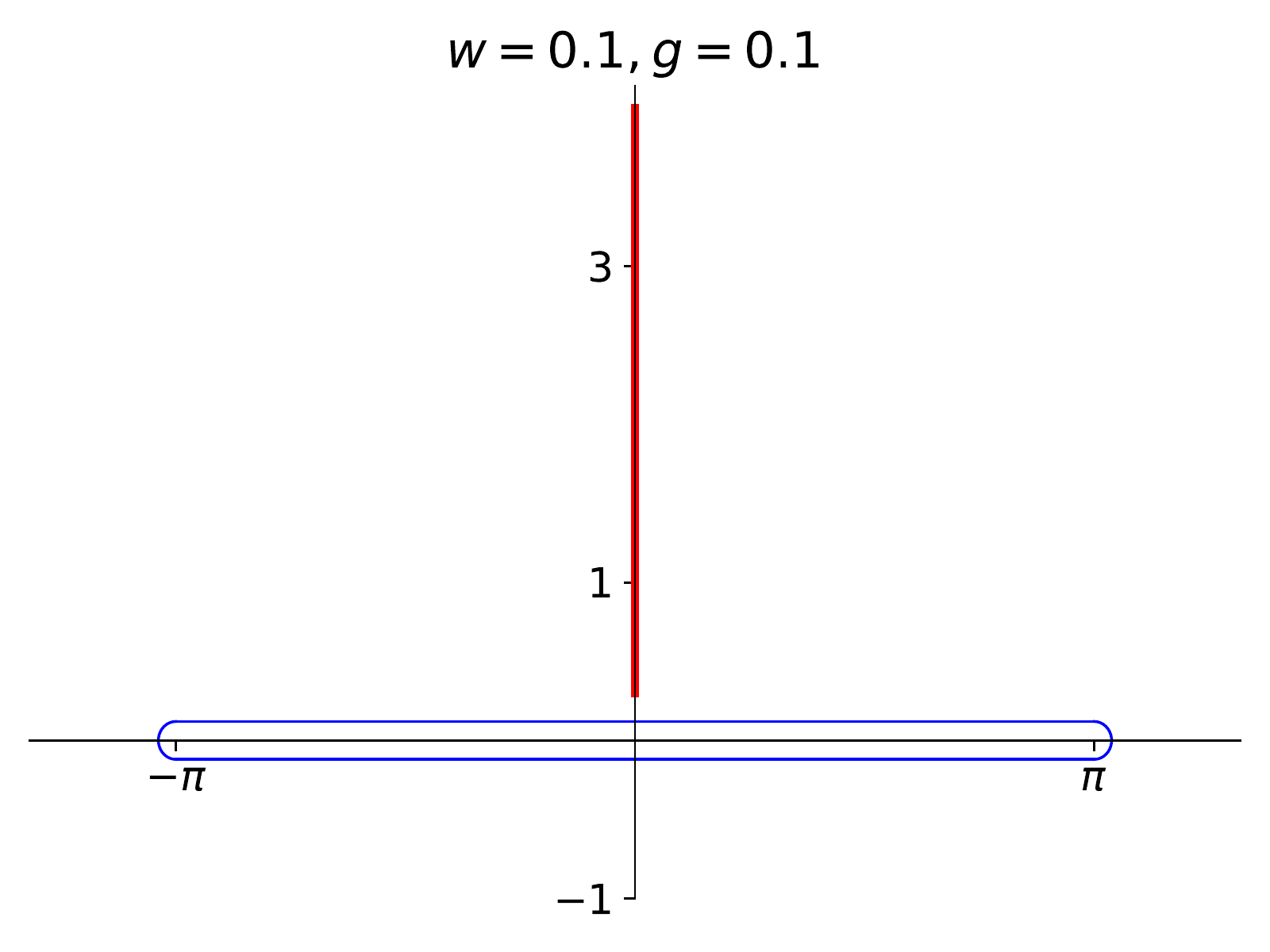}
            \centering
            \caption{(Color online) The (blue) integration contour , $C$, and the (red) branch cut at $\sqrt{z^2+w^2+8g^2}$ in the upper half-plane.
            }
            \label{contour}
        \end{figure}
        At $w<1$, corresponding 
    to the criticality and the $2$-MF phase, we find that 
the energy of bulk MFT-states, $|\psi_{n,k}\rangle$, and the edge-modes, $|\psi_{L,R}\rangle$, is sensitive to the scale $g$. Here we only present an expression for the g-dependence of the positive energy band, $\epsilon_l(g)$, which represents the energy of the MFT state with quantized $x_l$ and $\epsilon_{e}(g)$ is the energy of boundary $|\psi_{L,R}\rangle$:
        \begin{eqnarray}
        \epsilon_l(g)&=&\frac{1}{2N}\{\sqrt{w^2+x_l^2+8g^2\sin^2 x_l}+\sqrt{w^2+x_l^2}\}\\
        \epsilon_e(g)&=&\frac{1}{2N}\sum_{l\geq 0}(-1)^l\{\sqrt{w^2+x_l^2+8g^2\sin^2 x_l}-\sqrt{w^2+x_l^2}\} \label{ana_spe} \nonumber
        \end{eqnarray}
        where $x_l$ is defined and calculated in Appendix ~(\ref{Appendix}).
        
        When $w>1$, which corresponds to the 4-MF phase, the low-energy spectrum is given by
        \begin{eqnarray}
        \epsilon_e&\approx&\frac{1}{2N}\{\sqrt{w^2+x_0^2+8g^2w}-\sqrt{w^2+x_0^2}\}\\
        \epsilon_0&=&\frac{1}{2N}\{\sqrt{w^2+x_0^2+8g^2w}+\sqrt{w^2+x_0^2}\}\\
        \epsilon_l&=&\frac{1}{2N}\{\sqrt{w^2+x_l^2+8g^2\sin^2 x_l}+\sqrt{w^2+x_l^2}\}, l\geq 1 \nonumber
        \end{eqnarray} 
        where $\epsilon_0$ is the correction to the energy of the boundary mode of MFT states, and $\epsilon_l$ yields the correction to the bulk energy. 
        
        \section{Analytical derivation of the finite-size scaling function} \label{analytical_fini_app}
        At $|w|\ll 1$, we use Eq~(\ref{ana_spe}), which describes the energy levels of the system at $w\leq 1$, and rewrite the ground state energy as a sum of energy levels in Eq.~(\ref{gs}) as  
        \begin{eqnarray}
        E_G=&&\frac{1}{2N} \sum_{l\in \text{odd}}\sqrt{w^2+x_l^2}\nonumber\\
        &+&\frac{1}{2N} \sum_{l\in \text{even}}\sqrt{w^2+x_l^2+8g^2}.
            \end{eqnarray} 
       where we use the fact that at the region $|w|\ll 1$, $\sin^2 x_l=(x_l)^2/[w^2+(x_l)^2]\approx 1$ (recall that $x_l=\frac{\pi}{2}+l\pi$ when $w=0$).
        So, one needs to treat the quantized $x_l$ with odd $l$ and even $l$ seperately. To this end we define 
        \begin{eqnarray}
        q_\pm(k)=~\partial_k\ln(\cos\frac{1}{2}(Nk+\delta_m)\pm\frac{\pi}{4}),
            \end{eqnarray} 
        where the poles of $q_{+/-}$ are composed of quantized $x_l$ with even/odd $l$. 
        Now we can pick a contour, $C$, showed in Fig.(\ref{contour}), to encircle all the poles of $q_\pm$ on the real axis and avoid the branch-cut line of $E_{m,g}(z)=\sqrt{z^2+w^2+8g^2}$ in the complex $z$-plane. Then we can perform the contour integration as follows:
\begin{eqnarray}
E_G&=&-\frac{1}{4}\oint_C \frac{dk}{2\pi i}[E_{m,g}(k)q_+(k)+E_{m,0}(k)q_-(k)] \nonumber.
\end{eqnarray}
Then, we rewrite $q_s(k), s=\pm$ as
    \begin{eqnarray}    q_s(k)&=&\partial_k\ln \left( \sum_{n=\pm}\exp \left[ in\left(\frac{1}{2}(Nk+\delta_m)+s\frac{\pi}{4}\right) \right] \right) \nonumber
    \end{eqnarray} 
For the part of integration below(above) real line, we may need following expression of $q_s(k)$:
\begin{eqnarray}    q_s(k)&=&    \partial_k\ln \left( \exp \left[\pm i\left(\frac{1}{2}(Nk+\delta_m)+s\frac{\pi}{4}\right) \right] \right)\\
&+&\partial_k\ln \left(1+ \exp \left[ \mp i\left((Nk+\delta_m)+s\frac{\pi}{2}\right) \right] \right). \nonumber
    \end{eqnarray} 
In the first term above, the function "$\ln \exp$" yields identity function and then the contour integration of this term in Eq.(\ref{ge}) yields the bulk energy and boundary energy 
\begin{eqnarray}\nonumber
N\epsilon&=&\int_{-\pi}^{\pi}\frac{dk}{2\pi}[E(m,g)+E(m,0)]N/4\\
b&=&\int_{-\pi}^{\pi}\frac{dk}{2\pi}[E(m,g)+E(m,0)]\partial_k\delta_m/4.
\end{eqnarray}


		\bibliography{document.bib}

	\end{document}